\documentclass{solarphysics}
\usepackage[natbib,optionalrh]{spr-sola-addons} 
\usepackage{graphicx}                    
\usepackage[usenames]{color}                       
\usepackage{url}                         

\begin{document}


\begin{article}

\begin{opening}

\title{A Multi-Wavelength Analysis of Active Regions and Sunspots by Comparison
of Automatic Detection Algorithms}

%
\author{C.~\surname{Verbeeck}$^{1}$ \sep
        P.A.~\surname{Higgins}$^{2}$ \sep   
        T.~\surname{Colak}$^{3}$ \sep
        F.T.~\surname{Watson}$^{4}$ \sep   
        V.~\surname{Delouille}$^{1}$  \sep    
        B.~\surname{Mampaey}$^{1}$  \sep    
        R.~\surname{Qahwaji}$^{3}$
       }

%
\runningauthor{C. Verbeeck, \textit{et al.}}
\runningtitle{Multi-Wavelength Analysis of Active Regions and Sunspots}

%
  \institute{$^{1}$ Royal Observatory of Belgium, Belgium
                     email: \url{cis.verbeeck@oma.be}\\ 
             $^{2}$ Trinity College Dublin, Ireland
                     email: \url{pohuigin@gmail.com} \\
             $^{3}$ University of Bradford, UK
                     email: \url{t.colak@bradford.ac.uk} \\
             $^{4}$ University of Glasgow, UK
                     email: \url{f.watson@astro.gla.ac.uk}
             }

\begin{abstract}
Since the \emph{Solar Dynamics Observatory} (SDO) began recording $\sim$\,1~TB of data per day, there has been an increased need to automatically extract features and events for further analysis. Here we compare the overall detection performance, correlations between extracted properties, and usability for feature tracking of four solar feature-detection algorithms: the Solar Monitor Active Region Tracker (\textsf{SMART}) detects active regions in line-of-sight magnetograms; the Automated Solar Activity Prediction code (\textsf{ASAP}) detects sunspots and pores in white-light continuum images; the Sunspot Tracking And Recognition Algorithm (\textsf{STARA}) detects sunspots in white-light continuum images; the Spatial Possibilistic Clustering Algorithm (\textsf{SPoCA}) automatically segments solar EUV images into active regions (AR), coronal holes (CH) and quiet Sun (QS). One month of data from the SOHO/MDI and SOHO/EIT instruments during 12 May\,--\,23 June 2003 is analysed. The overall detection performance of each algorithm is benchmarked against National Oceanic and Atmospheric Administration (NOAA) and Solar Influences Data Analysis Centre (SIDC) catalogues using various feature properties such as total sunspot area, which shows good agreement, and the number of features detected, which shows poor agreement. Principal Component Analysis indicates a clear distinction between photospheric properties, which are highly correlated to the first component and account for 52.86\% of variability in the data set, and coronal properties, which are moderately correlated to both the first and second principal components. Finally, case studies of NOAA 10377 and 10365 are conducted to determine algorithm stability for tracking the evolution of individual features. We find that magnetic flux and total sunspot area are the best indicators of active-region emergence. Additionally, for NOAA 10365, it is shown that the onset of flaring occurs during both periods of magnetic-flux emergence and complexity development. 
\end{abstract}



\end{opening}

%
\section{Introduction}\label{intro}
%
In the 1960s, NASA launched the Pioneer 6, 7, 8, and 9 spacecraft, that were tasked with observing
the solar wind and interplanetary magnetic fields, forming the first space-based space-weather network and recording 512 bits per second. By comparison, the recently launched \emph{ Solar Dynamics Observatory} (SDO) is currently relaying solar data back to Earth at a rate of 150\,000\,000 bits per second.
With SDO returning the equivalent of an image with 4096 by 4096 pixels every
second, human analysis of every image would require a large team of people working
24 hours a day. The technological advances that have allowed the increased flow
of data, such as improving communication bandwidths and onboard processing
power, allows us to record data with a much greater cadence and spatial resolution than ever
before. However, there are problems with the storage, transfer, and analysis of such a large flow of data. SDO generates around 1~TB of data per day which is unprecedented in
solar physics. Getting this volume of data to researchers around the world,
as well as storing it in convenient places for analysis, is essential to make good use of it. An effective solution to the problem is to use automated feature-detection methods, which allow users to selectively acquire interesting portions of the full data set.

Development of automated solar feature detection and identification methods has increased dramatically in recent years due to the growing volume of data available. An overview of the fundamental image-processing techniques used in these algorithms is presented in \citet{Aschwanden_2010}. These techniques are used to detect many features in various types of observations at different heights in the solar atmosphere \citep{PerezSuarez2011}. In thie present work, we focus on detecting sunspot groups and active regions in photospheric continuum images, magnetograms, and EUV images. Previously, detection of sunspot groups in photospheric images was investigated by \cite{Zharkov2004} and \cite{Curto2008}. As well as detecting sunspot groups, \citet{Nguyen2005, Colak2008} also make automated classifications. The detection of active regions in magnetograms is explored in \citet{McAteer2005, labonte2007}, \citet{Lefebvre2004} and \citet{Qahwaji2006}, while \citet{Dudok06} introduces a supervised segmentation of EUV images into AR, CH, and QS regions.



The purpose of this article is to determine the robustness of four
algorithms for detecting and physically characterising active regions and sunspot groups by comparison of their outputs. We determine overall detection performance, the correlations between extracted feature properties using Principal Component Analysis, and the usability of these algorithms for tracking feature evolution over time. The tools that we consider are the Solar Monitor Active Region
Tracker \citep[\textsf{SMART}:][]{Higgins2010} which detects magnetic features using
magnetograms, the Automated Solar Activity Prediction code
\citep[\textsf{ASAP}:][]{Colak2009} which detects sunspots and pores using
photospheric intensity images, the Sunspot Tracking And Recognition Algorithm \citep[\textsf{STARA}:][]{Watson2009} which also detects sunspots in photospheric intensity images, and the
Spatial Possibilistic Clustering Algorithm \citep[\textsf{SPoCA}:][]{Barra2009} which
detects active regions in the corona using extreme ultraviolet images. More
detail on how these algorithms operate is provided in Section \ref{methods}. 

The overall performance of the algorithms is compared by determining the total number of features detected as well as their full-disk area. Our methods are benchmarked against National Oceanic and Atmospheric Administration (NOAA) and Solar Influences Data Analysis Centre (SIDC) catalogues. Few studies include comparisons of detection methods or different data types. \citet{Benkhalil2006} compare the detection of active regions in several data types using a single region-growing method. Other studies compare the detection of features in magnetograms using a variety of region-growing and morphological methods \citep{deforest2007,parnell2009}. Direct comparison of algorithms is important for their characterisation, as each is designed in a different way to detect features for a specific purpose. 

Correlations between the properties determined by the algorithms are investigated using Principal Component Analysis~\citep{Jolliffe2002}. \textsf{PCA} has been used previously for various purposes in solar-physics and space-weather literature, \emph{e.g.}  to detect outliers~\citep{2011A&A...528A..62S}, to reduce dimensionality~\citep{2007A&A...466..347D}, or for exploratory data analysis of space-weather data sets~\citep{2011AdSpR..47.2199H}.

Finally, the stability of the algorithms is tested for tracking feature evolution through time. The evolution of two ARs is studied in detail, including their emergence
in several layers of the solar atmosphere. \citet{Lites1995} present a similar
multi-layered analysis of the emergence of an AR. As non-potentiality increases
in
an AR, it may begin to exhibit enhanced coronal activity. This effect has been
studied in many articles, and it is related to dynamic behaviours such as helicity
injection \citep{MoritaMcintosh_2005}, turbulent cascades
\citep{hewett_2008,conlon_2008,conlon_2010}, enhanced polarity separation line gradient
\citep{falconer_2008}, and changes in magnetic connectivity
\citep{georgoulis_2007,Ahmed2010}. In this article we study
multiple behaviours in the same AR using magnetic property determinations.
Finally, the decay of the AR in the corona and photosphere is compared. To
our knowledge, this is the first time that automated feature-detection algorithms have been used to study 
temporal evolution using properties of magnetic non-potentiality, sunspot
characteristics, and coronal activity of ARs simultaneously.

The following sections detail these investigations. 
Observations used in this study are described in Section \ref{datasets}, and the four algorithms to be compared are introduced in Section   \ref{methods}. 
 Our results  are presented in Section
\ref{results}, including an evaluation of the algorithms' overall performance, a correlation study
of the complete sample of active regions and a detailed case study of two different active regions.
Finally, a discussion of the results and concluding remarks is presented in Section \ref{disc_concl}.

\section{Observations}\label{datasets}

In this study we analyse data from the interval 12 May\,--\,23 June 2003.
The detections obtained from each algorithm for the entire data set are
studied as a whole in Section \ref{correlation} and NOAA ARs 10377 and 10365 are
individually studied in detail in Section \ref{case_studies}. This
particular data set was selected for the diversity of solar features present. 
SOHO/MDI magnetograms are used for magnetic region detection by \textsf{SMART},
while SOHO/MDI continuum images are used for sunspot detection by
\textsf{ASAP} and \textsf{STARA}, and SOHO/EIT images are employed for active region
detection by \textsf{SPoCA}. These algorithms are described in Section \ref{methods}.

The MDI instrument on SOHO provides almost continuous observations of the Sun in
the white-light continuum, in the vicinity of the Ni\,{\sc i} 676.78 nm photospheric
absorption line. These photospheric intensity images are primarily used for sunspot observations.
MDI data are available in several processed ``levels''. We used level-2 images, which are smoothed, filtered, and rotated
\citep{Scherrer1995}.  SOHO provides two to four MDI photospheric intensity images per
day with continuous coverage since 1995.

Using the same instrument level, 1.8 line-of-sight (LOS) MDI magnetograms are
recorded with a nominal 
cadence of 96 minutes. The magnetograms show the
magnetic fields of the solar photosphere, with negative (represented as black)
and positive (as white) areas indicating opposite LOS magnetic-field
orientations.

The \emph{ Extreme Ultraviolet Telescope}
(EIT: \opencite{1995SoPh..162..291D}),
onboard SOHO,
delivers synoptic observations consisting of 1024 by 1024 images of the solar corona
recorded in four different wavelengths every six hours.
Every \textsf{SPoCA} segmentation in this article was based on a pair of 17.1 and 19.5~nm
EIT images. All images used have been preprocessed using the standard \textsf{eit-prep} procedure of the \textsf{SolarSoftware} library. A fixed-centre
segmentation with six classes was performed on the logarithms of
the image pixel values. The AR centre values are 401.74 and 324.25
DN\,s$^{-1}$ in 17.1 and 19.5 nm, respectively. These values were derived from a
cumulative run
of \textsf{SPoCA} on a data set of monthly EIT image pairs from February 1997 until April
2005; see~\citet{Barra2009}.

In the case studies presented in Section \ref{case_studies}, we compare
observations of NOAA 10365 and 10377 with flares characterised by the
\emph{Reuven Ramaty High Energy Solar Spectroscopic Imager} (RHESSI) team and
distributed in the RHESSI flare
list.

\section{Methods}\label{methods}

\textsf{SMART}, \textsf{ASAP}, and \textsf{STARA} detect photospheric features such
as active regions and sunspots while \textsf{SPoCA} uses images in the extreme ultraviolet
to observe active regions at the coronal level. 
In this section, we will describe each of the feature-detection algorithms used
(Sections \ref{smartsect}, \ref{asapsect}, \ref{starasect}, and
\ref{spocasect}). The outputs from each of the algorithms are associated using
the method explained in Section \ref{association_algorithm}.

\subsection{The \textsf{SMART} algorithm}\label{smartsect}

\begin{figure}
\centerline{\includegraphics[width=0.9\textwidth,clip=]{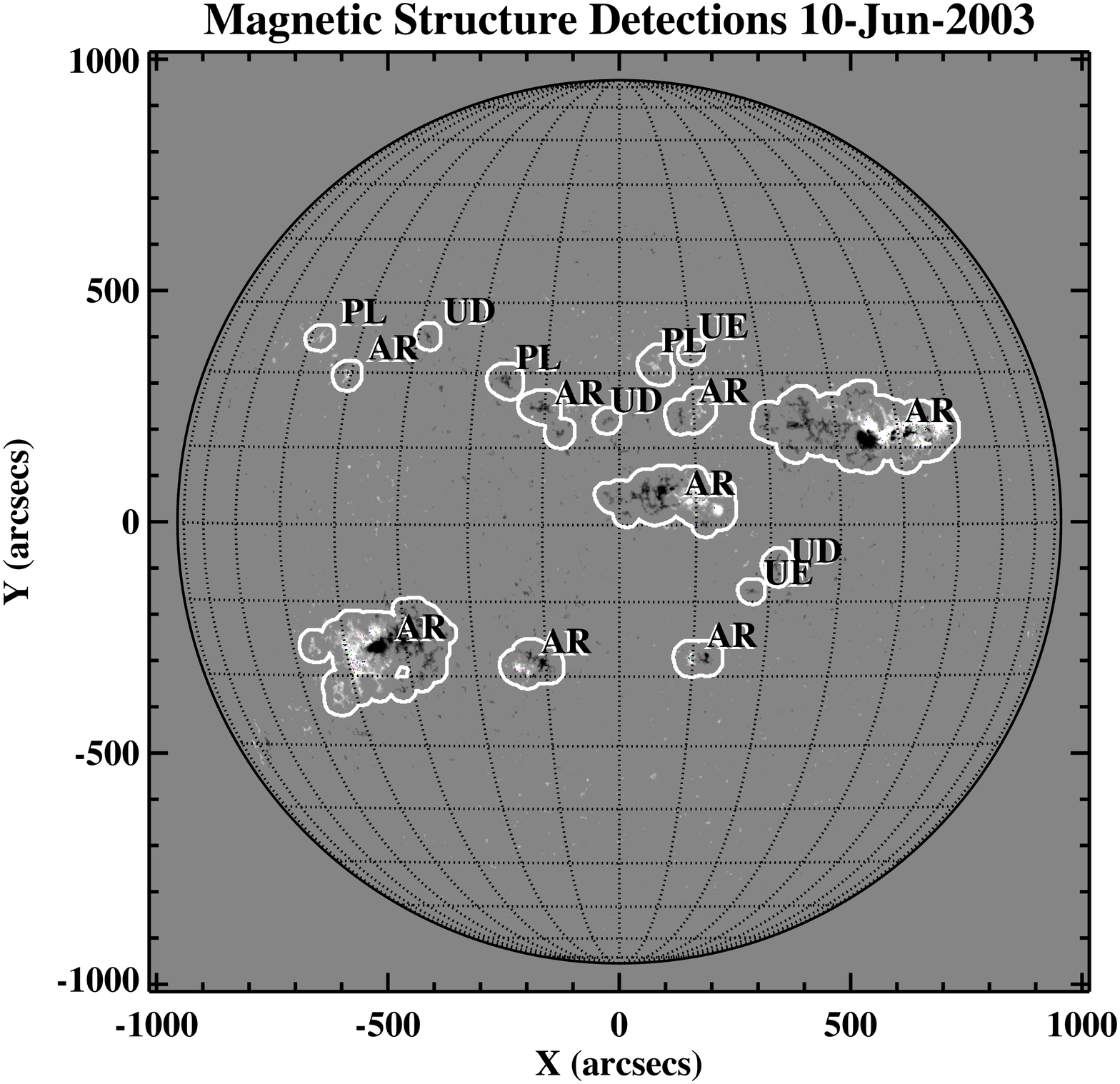}}
\caption{An example set of \textsf{SMART} detections from 10 June 2003. PL, UD, and UE
identify three classes of unipolar feature, while AR denotes multipolar
features.}\label{smartexampdetect}
\end{figure}

The Solar Monitor Active Region Tracker \citep[\textsf{SMART}:][]{Higgins2010} is an algorithm that uses
magnetograms to
automatically extract, characterise, and track active regions over multiple
solar rotations -- from first emergence to decay.
This allows one to study the complete life-cycle of ARs. The algorithm uses a combination
of image-processing techniques to determine the boundary of an AR.
Two consecutive line-of-sight magnetograms are smoothed using a gaussian kernel with
a full-width at half-maximum of five~pixels and thresholded by 70~G to identify
potential features. The two detections are overlaid to identify and remove
features that are not present in both magnetograms. The remaining detection
boundaries are then dilated by ten pixels to create the final mask.

Dilation is
performed to include nearby decaying and plage fragments that may have separated from the
main AR. This is intended to help conserve the measured polarity balance of the
AR as it evolves. An example set of \textsf{SMART} detections is shown in Figure
\ref{smartexampdetect}. In this article, SOHO/MDI LOS magnetograms are used
for detection, but recently the algorithm has been adapted for use with
SDO/HMI magnetograms (For near-realtime detections, see
\url{http://solarmonitor.org/smart_disk}).

\begin{figure}
\centerline{\includegraphics[width=0.9\textwidth,clip=]{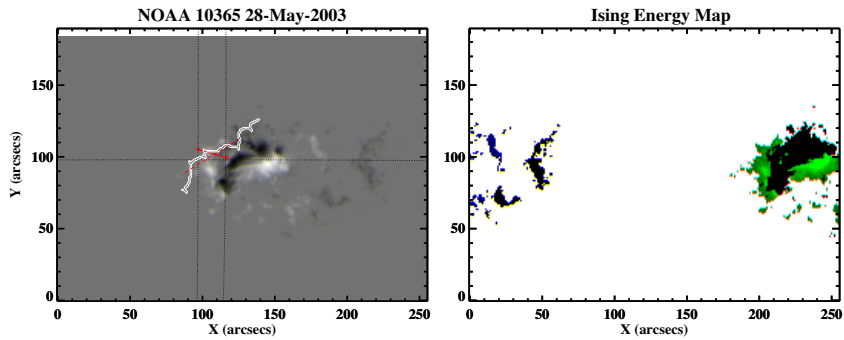}}
\caption{Left: NOAA 10365 highlighting the PSL (white contour), a linear fit to the locus of PSL positions (dotted red line), bipolar
connection line (solid red line),
and heliographic longitude and latitude reference lines (dotted black lines). Right: An
Ising energy map of
the same region. Red represents the magnitude of energy for each pixel from
highest (light) to lowest (dark). Since the connection between pixels of
opposite polarity is being represented, the energy map is only shown for one of
the polarities.}\label{smartbipolepslexamp}
\end{figure}

Several new physical-property modules have been added to \textsf{SMART} for this study.
The tilt of an AR is obtained by measuring the angle between
the line connecting the centroids of the largest flux-weighted positive and negative blobs (solid red line in left panel of Figure \ref{smartbipolepslexamp}) and the heliographic-latitude line passing through the
centroid of the AR. The length of this bipolar connecting line [BCL] line is also 
determined and provides a measure
of the relative compactness of an AR when compared to the evolution of AR area 
(left panel of Figure \ref{smartbipolepslexamp}). Additionally, the angle [$\alpha$]
detected between the best-fit line to the locus of pixels forming the main polarity separation line [PSL] and the BCL is measured. The main PSL is defined as the interface between the aforementioned largest flux-weighted positive and negative blobs.
The temporal derivative of the angle $\alpha$ is shown to be a useful proxy for the occurrence of helicity
injection in an AR \citep{MoritaMcintosh_2005}, which may be an
important flare predictor \citep{labonte_2007}. The evolution of this angle is studied in Section \ref{case_studies}. These properties are less
informative when studying AR complexes or non-bipolar
ARs, since often no main axes can be discerned, making a description of the AR
orientation impossible.

The original Ising model is used for the analysis of magnetic interactions and
structures of ferromagnetic substances. Here, as by \citet{Ahmed2010}, Ising energy is used as a proxy for
magnetic connectivity and complexity within an AR (right panel of Figure \ref{smartbipolepslexamp}). We use a modified form of that given by \citet{Ahmed2010}.
 
This version is calculated using
\begin{equation}
\displaystyle\sum\limits_{i,j} \left|\frac{B_i B_j}{D_{ij}}\right| \mbox{,} 
\end{equation}
where $B_i$ ($B_j$) are pixel values of positive (negative) line-of-sight magnetic field and $D_{ij}$ is the spatial distance between pixels $i$ and $j$.   
The Ising energy increases as the negative and positive magnetic footprints
within an active region become more entangled. The evolution of this property is studied in Section \ref{case_studies}.

The modules added to \textsf{SMART} for this work will be used for future large-scale active region studies and will be added to the pipe-line versions of \textsf{SMART} running at the Heliophysics Events Knowledgebase
\citep[\url{http://www.lmsal.com/hek/index.html}:][]{hurlburt_etal_2010} and included in the Heliophysics Integrated
Observatory \citep[\url{http://www.helio-vo.eu/index.php}:][]{Bentley_2009}. In the future, other property modules will be added to
calculate a physically motivated magnetic-connectivity measurement
\citep{georgoulis_2007}, multi-scale energy spectrum slope
\citep{hewett_2008}, and multi-fractal spectrum properties \citep{conlon_2008}.

\subsection{The \textsf{ASAP} algorithm}\label{asapsect}

\begin{figure}
\centerline{\includegraphics[width=0.9\textwidth,clip=]{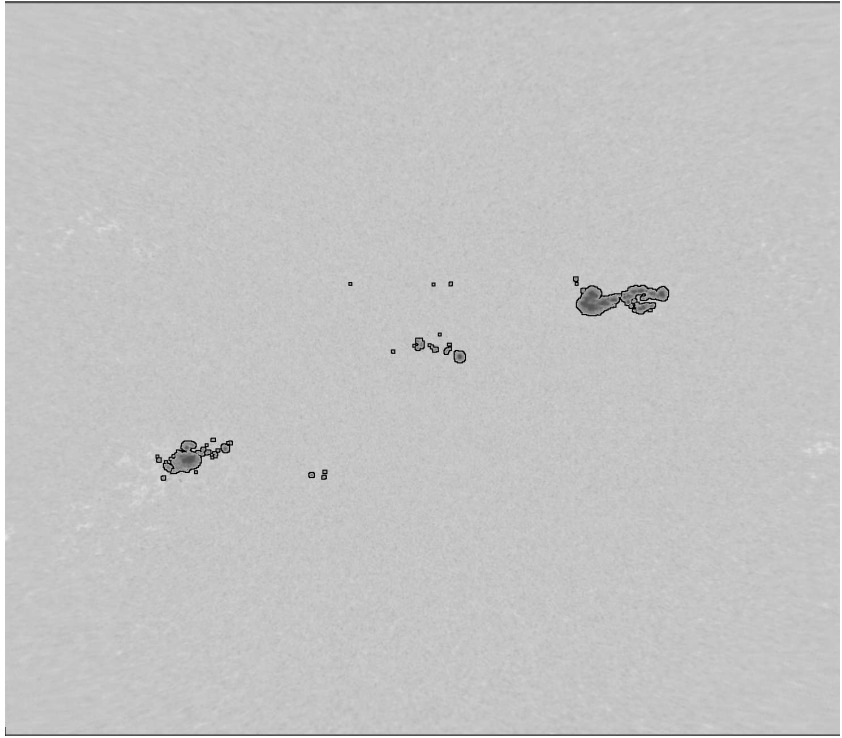}}
\caption{An example set of \textsf{ASAP} detections from 10 June
2003.}\label{asapexampdetect}
\end{figure}

Automated Solar Activity Prediction (\textsf{ASAP}) is the collective name for a set of
algorithms used to process solar images. It is composed of algorithms for
sunspot, faculae, and active-region detections \citep{Colak2008} and solar-flare
prediction \citep{Colak2009}. Unlike other algorithms described in this article,
\textsf{ASAP} uses quick look (in GIF or JPEG format) images for its processes. In
this article a recently developed sunspot detection algorithm for \textsf{ASAP} is used.
This new sunspot-detection algorithm works with continuum images and is
described in detail by \cite{Colak2010}. The main steps in this algorithm can
be summarized as follows:

\begin{itemize}
\item Images are pre-processed to detect the solar disk, and to remove limb
darkening.
\item Detected solar-disk data are converted from heliocentric coordinates to
Carrington heliographic coordinates.

\begin{itemize}
\item The key point in heliographic conversion is to choose the size of the
resulting image. If a very small image size is selected, this will cause
truncation and loss of data. If a very large one is chosen, there will be many
spaces in the resulting image. In this study, each heliographic degree is
represented by ten pixels therefore the resulting heliographic images are 3600 by
1800 pixels.
\item Initially, an empty 3600 by 1800 image is created. When the Carrington longitude and latitude of all of the pixels on the solar disk are
calculated, their pixel intensities are placed in corresponding locations.
\item The distribution of pixels representing degrees in heliocentric coordinates is not uniform due to the spherical shape of the Sun. Towards the limb of the Sun on a two dimensional heliocentric image each degree will be represented by fewer pixels although in a heliographic image each degree is represented with same amount of pixels. Size of the heliographic image is larger than heliocentric image and therefore there will be gaps (empty pixels that are not filled with data from heliocentric image) in the resulting initial heliographic image after conversion.
\item The following algorithm is applied to estimate the pixel intensities of the gaps in the heliographic image. Every pixel on the heliographic image is examined and when a pixel without any information is found, its neighbouring pixels are searched using variable-size windows. First a 3 × 3 pixel window is centred on the empty pixel and the values of the non-empty neighbouring pixels within this window are added. If all the neighbouring pixels within the initial window were empty, the size of the window is increased by one and the process continued until at least one pixel with information is available within the window. Then the average value of the valid pixels found within the final window is assigned to the empty pixel. The algorithm continues until all of the pixels have been processed.
\item After all data gaps have been filled, a smoothing algorithm using a $3\times3$ linear uniform filter is applied to create the final heliographic
image.
\end{itemize}
\item Subsequently, the following filter is applied to the Carrington heliographic image for detecting
sunspots:
\begin{itemize}
\item An intensity filtering threshold value $T = \mu - (\sigma \times \alpha)$ is calculated where $\mu$ is the
mean, $\sigma$ is the standard deviation of the image, and $\alpha$ is a
constant equal to 2.5. 
\item The intensity of each pixel in the image is compared to this $T$
value. If it is less than the calculated threshold value, the pixel under
consideration is marked as a sunspot.
\end{itemize}
\end{itemize}

Although heliographic conversion can be computationally expensive, it yields detections that are more accurate compared to the ones done on heliocentric
coordinates. A tracking algorithm was added to \textsf{ASAP} for this study, finding the
intersections between objects (\emph{e.g.} sunspots) on two consecutive heliographic
images. Since the differential rotation is very small in Carrington heliographic
coordinates, there is no need for longitudinal corrections. \textsf{ASAP} tends to detect small sunspots, which can be classified as
pores. This is useful especially when grouping and classifying sunspots. However, the
number of tracked sunspots increases because most pores are only visible for a few hours on the solar disk.
An example set of \textsf{ASAP} detections is shown in Figure~\ref{asapexampdetect}.

It was not necessary to update \textsf{ASAP} for this work, but some computational issues
have been discovered. For instance, the application of \textsf{ASAP} to SDO/HMI images larger than $1024\times1024$~pixels shows that heliographic conversion algorithms must be made more efficient to tackle larger images.

\subsection{The \textsf{STARA} algorithm}\label{starasect}

\begin{figure}
\centerline{\includegraphics[width=0.9\textwidth,clip=]{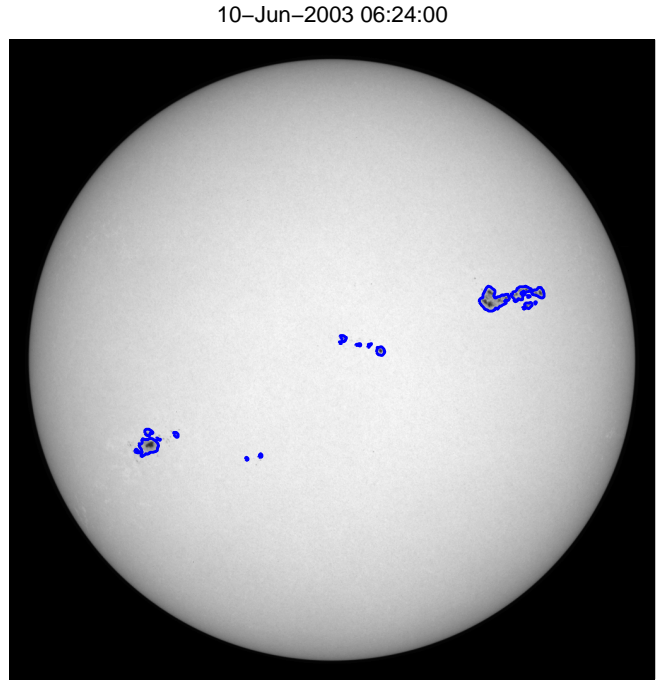}}
\caption{An example set of \textsf{STARA} detections from 10 June
2003.}\label{staraexampdetect}
\end{figure}

The Sunspot Tracking And Recognition Algorithm (\textsf{STARA}) code was written in 2008
in order to perform consistent long term observations of sunspots over Solar
Cycle 23 \citep{Watson2009}. It was originally developed for use with MDI
data but has since been extended for use with data from SDO as well as from a number of ground-based instruments. A simple
detection method was required to speed up processing when large data sets were
used and suitable techniques were found in the field of morphological image
processing.

The \textsf{STARA} detection method works as follows: The image is read in and
inverted so that the sunspots appear as bright areas on a dark background for compatibility with the
top-hat transform. Morphological erosion is applied, which removes peaks and works by treating the 2D image as a 3D surface with the pixel values indicating the height of the surface at that
point. A probe known as a structuring element is then chosen (in this case, a
sphere with a radius of 14 MDI pixels) and is ``rolled'' underneath the
surface whilst always touching it. The centre of the sphere then maps out a new
surface that is close to 14 units below the original. However, any sharp peaks present
(such as sunspots) would not allow the sphere to fit inside them and so
are not represented in the eroded image. A morphological dilation is then
performed which is identical to an erosion apart from the sphere being rolled on
top of the surface. The dilated surface is subtracted from the original to leave only the sunspot peaks present.

As the sphere rolls over the surface it also carries the limb-darkening profile through each step and when the final surface is subtracted from the original, the limb-darkening effects are automatically removed.

More detail on this step and the top-hat transform is given
by \citet{Watson2009} and \citet{Dougherty2003}. A size filter is then applied
that removes areas containing less than ten pixels as they are far more likely to be pores than sunspots. The remaining areas are recorded
along with their locations as well as the number of umbral regions detected
within the sunspot boundary. This is repeated for a number of consecutive images
and using the solar-rotation model of \citet{Howard1990} the sunspots can be
tracked throughout a sequential data set. This allows the evolution of individual
sunspots to be followed as well as the overall properties of the sunspot
population as a whole. An example of a typical set of \textsf{STARA} sunspot detections
is given in Figure~\ref{staraexampdetect}.

\textsf{STARA} has undergone very few changes over the course of this work as the code
was well established beforehand. Nevertheless, some subtle problems have been
discovered in the process.
As sunspots approach the limb (at longitudes greater than
75$^{\circ}$) the sunspot position returned by the code quickly loses accuracy.
This is a common problem with feature-detection methods as the geometrical
foreshortening effects test the limits of automated systems. There are also
potential problems present with bad data. Obviously the best remedy is to remove
it altogether but with MDI it is possible to have images with only half of
the solar disk present or with large artifacts. Both of these situations can have
substantial effects on the detected global properties and cause problems with analysis.

\subsection{The \textsf{SPoCA} algorithm}\label{spocasect}

\begin{figure}
\centerline{\includegraphics[width=0.9\textwidth,clip=]{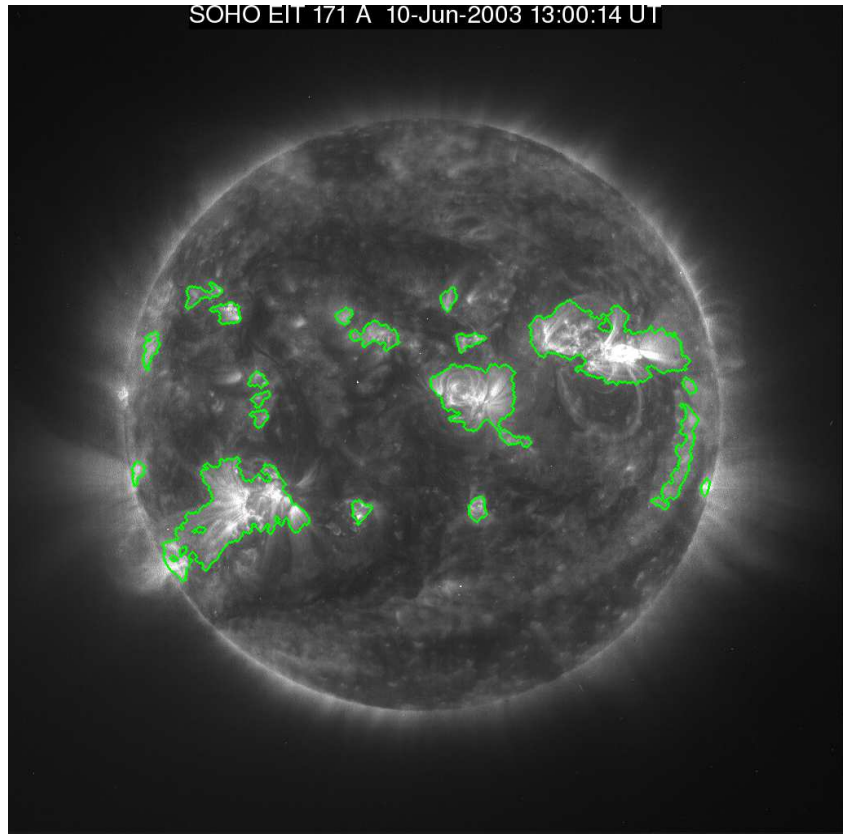}}
\caption{An example set of \textsf{SPoCA} detections from 10 June
2003.}\label{spocaexampdetect}
\end{figure}

The Spatial Possibilistic Clustering
Algorithm (\textsf{SPoCA}) is a multi-channel fuzzy clustering algorithm that automatically segments solar
EUV images into a set of features; see  \citet{Barra2009} for a complete presentation. It optimally separates active
regions, quiet Sun, and coronal holes, even though the boundaries
of these regions are not always well defined. The description of the
segmentation process in terms of fuzzy logic was motivated by the facts that
information provided by a solar EUV image is noisy (corruption by Poisson and
readout
noise as well as by cosmic-ray hits) and subject to both observational biases
(line-of-sight
integration of a transparent volume) and interpretation (the apparent boundary
between regions is a matter of convention).

\textsf{SPoCA} takes as input an image in one (or several) EUV passband(s) and uses as ``feature vector''
the pixel value (or the pixel value vector in the multi-channel case) in order to classify a pixel as belonging to one of three classes, namely AR, QS, and CH. 
\textsf{SPoCA} is based on  a fuzzy clustering technique called ``Possibilistic C-Means''
\citep[\textsf{PCM}:][]{Krishnapuram93,Krishnapuram96}. For each class, it assigns a ``probability'' or membership value $\in[0,1]$ to every feature vector.

\textsf{PCM} is an iterative method that searches for three compact clusters in the space
of feature vectors, corresponding to AR, QS and CH.
In practice, this is achieved through a gradient-descent scheme that minimizes an
objective function that is related to the total intracluster variance plus some penalty term. In every iteration, new membership values are calculated based on the class centre values. The membership values are used in turn to compute the new class centres, and so on, until the
class centres converge to within a preset accuracy.

In order to cope successfully with intensity outlier pixels such as those
affected by cosmic rays and proton storms, a spatial regularization term was
added to the \textsf{PCM} objective function, forcing membership values in a neighbourhood
to be as close as possible. By assigning each pixel to the class for which its feature vector has the largest membership value, the image is segmented.
An example set of \textsf{SPoCA} detections is shown in Figure~\ref{spocaexampdetect}.

Since the solar corona is optically thin, and since
the intensity in EUV images is obtained through an integration along the line of
sight, there is a limb-brightening effect in those images, which may hinder the
segmentation process. Therefore, the EUV images are pre-processed so as to lower
the enhanced brightness near the limb. The initial \textsf{SPoCA} class contours are
automatically postprocessed using a morphological opening with a
circular isotropic element of size unity.

Since the publication of~\citet{Barra2009}, the \textsf{SPoCA} algorithm was
optimized and extended in several ways:
\begin{itemize}
\item
In order to gain more consistent results, we introduced some constraints on the penalty term of the objective function to be minimized.
\item
The limb correction is now applied in a continuously increasing way towards the
limb instead of introducing it abruptly from some point onwards.
\item
For individual AR detection, first the Bright Points are removed (size threshold is 1500 square arcseconds) and then a spherical dilation (radius: 12 EIT pixels) is employed to group the
remaining bright blobs into individual active regions.
\item
Individual AR are tracked through time by comparing the masks of regions in two
consecutive time frames, taking into account differential rotation.
\end{itemize}

\textsf{SPoCA} has been running in near-realtime on AIA data since September 2010 as part
of the SDO Feature Finding Project~\citep{Martens_etal_2011}, a suite of software pipeline modules for
automated feature recognition and analysis for the imagery from SDO. The resulting AR events are automatically
ingested by the Heliophysics Events Knowledgebase
\citep{hurlburt_etal_2010}.

\textsf{SPoCA} is the only algorithm presented here that detects ARs in the solar
corona.
The method is generic enough to allow the introduction of other channels or
data. It has been applied to SOHO/EIT, SDO/AIA, PROBA2/SWAP,
and STEREO/EUVI images, and could
potentially be used on other multi-channel maps such as Differential Emission
maps. In this article we focus on ARs, but QS and CH can also be detected and
tracked.

\subsection{Association of Detected Features}\label{association_algorithm}

The \textsf{SMART} tracking module,  called ``Multiple Disk Passage'' \citep[\textsf{MuDPie}:][]{Higgins2010}, is used to associate
individual \textsf{SMART} detections of the same physical feature over time by comparing the
centroids of all detections in consecutive magnetograms. Two detections are
associated if their centroids match within $5^{\circ}$ heliographic latitude and
longitude. The tracked \textsf{SMART} detections are then associated with
the best matched detections in each of the other algorithms as described in the
following paragraphs.

In order to analyse the relation between the features detected by different
algorithms, a routine developed in Python associates detections
from each algorithm in two ways. First outputs from \textsf{ASAP}, \textsf{STARA}, and
\textsf{SPoCA} are associated with \textsf{SMART} outputs based on time and location
information. Second, individual association outputs (\textsf{ASAP} \emph{vs.} \textsf{SMART}, \textsf{STARA} \emph{vs.}
\textsf{SMART}, \textsf{SPoCA} \emph{vs.} \textsf{SMART}) from the first step are combined using \textsf{SMART} IDs and
timing information.

For associations, \textsf{SMART} is chosen as the base algorithm because \textsf{SMART} detections
usually encircle the corresponding \textsf{ASAP} and \textsf{STARA} detections and they are also more
stable over time than \textsf{SPoCA} detections, due to the frequent splitting and
merging of coronal AR detected by \textsf{SPoCA}. 
Also, \textsf{SMART} detects magnetic regions from MDI images
which are more frequently available than the continuum and EIT images that the
other algorithms are working on. The association rules are described below.

\medskip

First Step: Individual associations (\textsf{ASAP}, \textsf{STARA}, \textsf{SPoCA} \emph{vs.} \textsf{SMART})

\begin{itemize}
\item{The time difference between the solar detections under consideration
(\emph{i.e.}, sunspots from \textsf{ASAP} and \textsf{STARA}, active regions from \textsf{SPoCA} \emph{versus} magnetic
regions
from \textsf{SMART}) is calculated.}
\item{If the time difference between a magnetic region detected by \textsf{SMART} and a
solar region detected by another algorithm is less than 0.2 Julian days and their
heliographic bounding boxes intersect, then these detections are associated.
Since \textsf{SPoCA} does not deliver heliographic bounding boxes, a bounding box of 5$^{\circ}$ in longitude and latitude is assumed.}
\item{If the same solar detection is associated to more than one \textsf{SMART} region,
only the closest (in terms of time and distance between centres) \textsf{SMART} region is
selected as associated.}
\item{Associations are saved in separate files (three files; \textsf{ASAP} \emph{vs.} \textsf{SMART}, \textsf{STARA} \emph{vs.}
\textsf{SMART}, \textsf{SPoCA} \emph{vs.} \textsf{SMART}) including the selected characteristics from each
algorithm.

output that is going to be analysed.}
\end{itemize}

Second Step: Combining all of the associations

\begin{itemize}
\item{The \textsf{SMART} algorithm uses an ID for each magnetic region detected and in
this second step, the association data saved in the three separate files from the
first step are combined using this ID and time information. The association data
with same \textsf{SMART} ID and closest timing are combined together. Timing information
still has to be used due to the difference between the image times.}
\item{The final combined data are saved in one file.}
\end{itemize}

\textsf{SMART} provided 9356 detections (207 magnetic region features), \textsf{ASAP} 3039
detections (952 sunspot features), \textsf{STARA} 1329 detections (433 sunspot features)
and \textsf{SPoCA} 1222 detections (190 coronal active-region features) within the
considered time-frame. In the first step, 714 \textsf{SMART} detections were associated to
2889 \textsf{ASAP}
detections, 550 \textsf{SMART} detections were associated to 1315 \textsf{STARA} detections and
1089 \textsf{SMART} detections were associated
to 1117 \textsf{SPoCA} detections. In
the second step when all of these data were combined, 350 detection (33 feature)
associations were created for \textsf{SMART}, \textsf{ASAP}, \textsf{STARA} and \textsf{SPoCA}.
The daily averages of some of the outputs such as average daily sunspot numbers,
active region numbers and average areas are compared to the NOAA active region
catalogue in Section~\ref{performance}. In the considered period, NOAA recorded
217 detections (37 features).

In case of merging or splitting of neighbouring coronal regions as detected by \textsf{SPoCA}, the association
procedure described above does not relate the new \textsf{SPoCA} detection to the
corresponding \textsf{SMART} detection. This happened 
several times 
in the case
studies in
Section~\ref{case_studies}. For these cases, we applied a manual association of
\textsf{SPoCA} detections to \textsf{SMART} detections.

\section{Results}\label{results}

The feature detections from each algorithm are compared in the following
sections. First, in Section~\ref{performance} the overall detection performance
of the
algorithms is presented, and compared to the corresponding NOAA
detections and the daily international sunspot number.
Next, Principal Component Analysis is performed on the full set of detections to
probe the overall structure of the physical properties calculated by the
algorithms in Section \ref{correlation}.
Finally, in Section~\ref{case_studies} the evolution
and flare activity of NOAA 10377 and 10365 are analysed in depth, using physical
properties determined by each algorithm.

\subsection{Algorithm Performance}\label{performance}

The performance of the algorithms is measured by comparing the daily
total and average values of some of the solar feature properties to each other,
to values reported by
NOAA (\url{http://www.swpc.noaa.gov/ftpdir/forecasts/SRS/README}), and to
the international daily sunspot numbers \citep{SIDC} between 12 May and 23 June 2003.

\begin{figure}
 \centerline{
 \hspace*{0.015\textwidth}
 \includegraphics[width=0.45\textwidth,clip=]{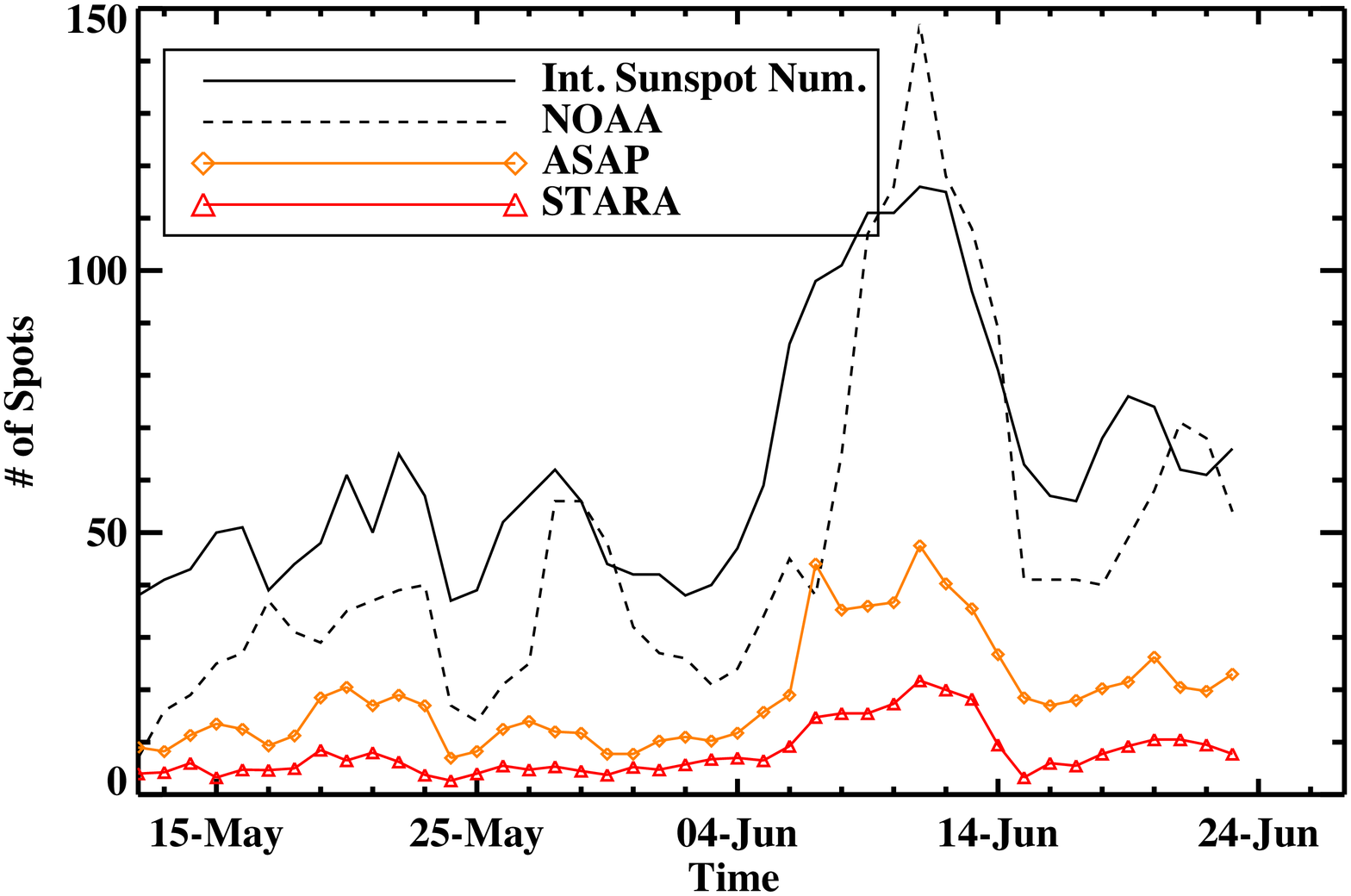}
   \hspace*{0.030\textwidth}
 \includegraphics[width=0.45\textwidth,clip=]{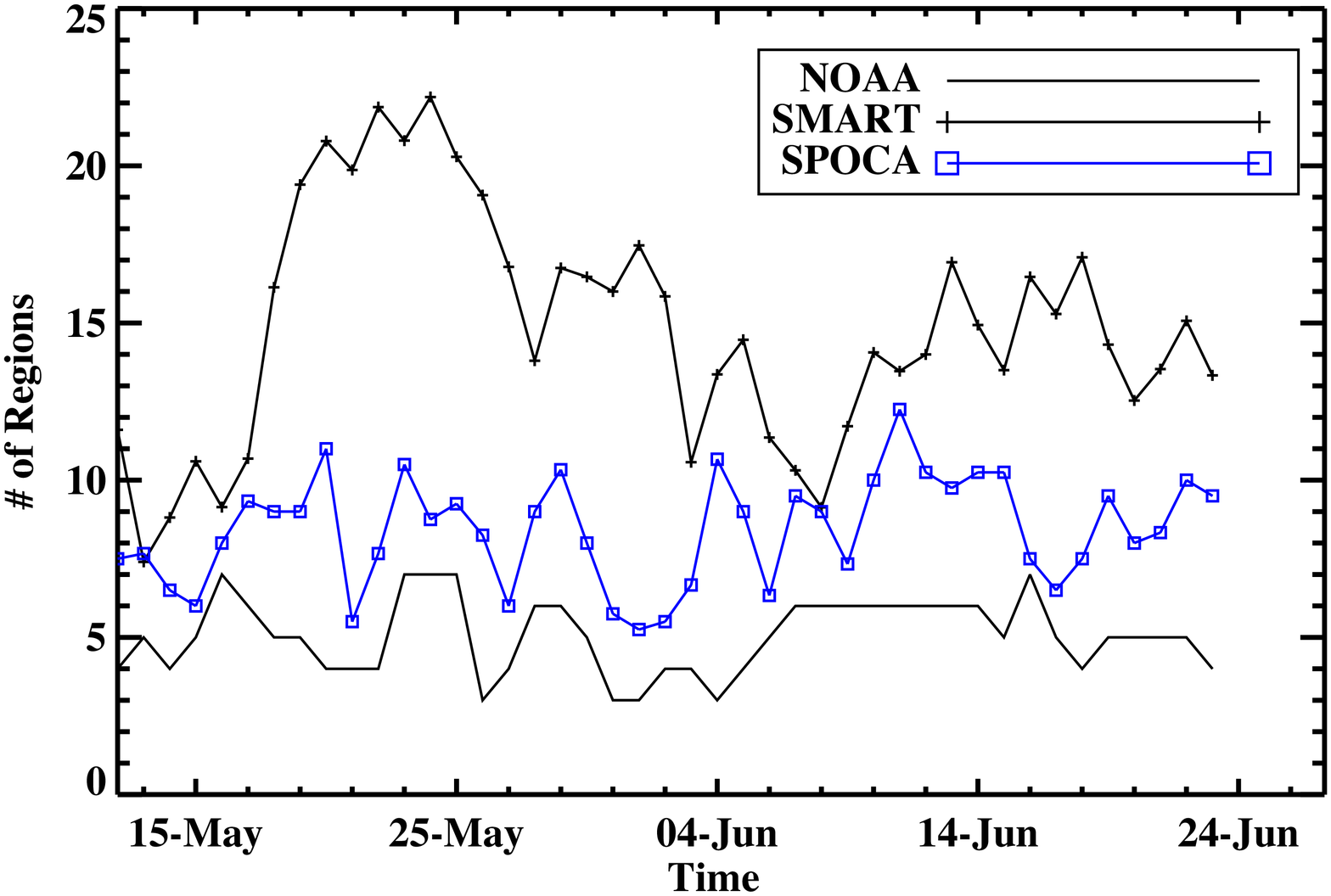}}
     \centerline{
 \hspace*{0.015\textwidth} 
 \includegraphics[width=0.45\textwidth,clip=]{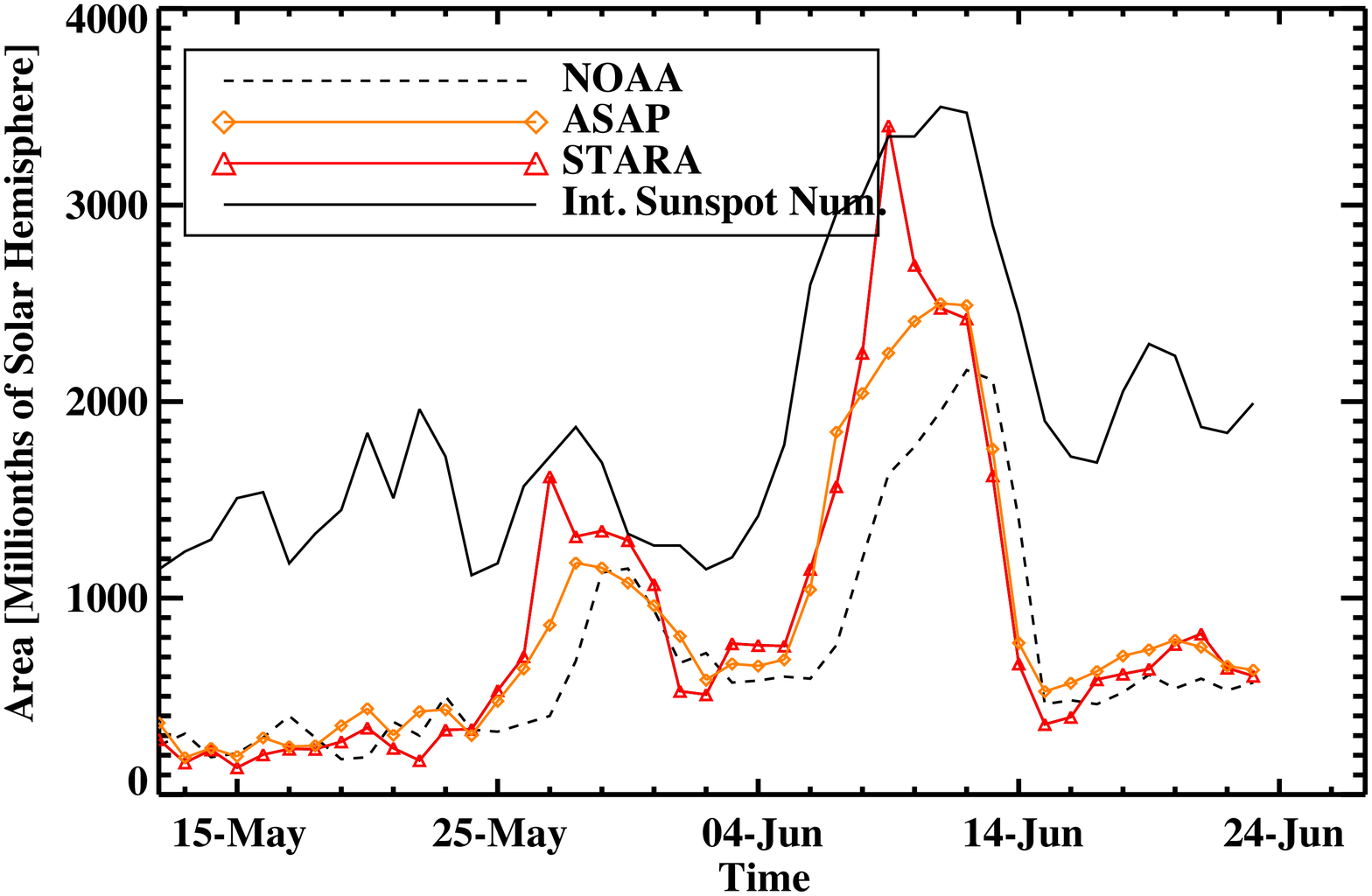} 
   \hspace*{0.030\textwidth}
\includegraphics[width=0.45\textwidth,clip=]{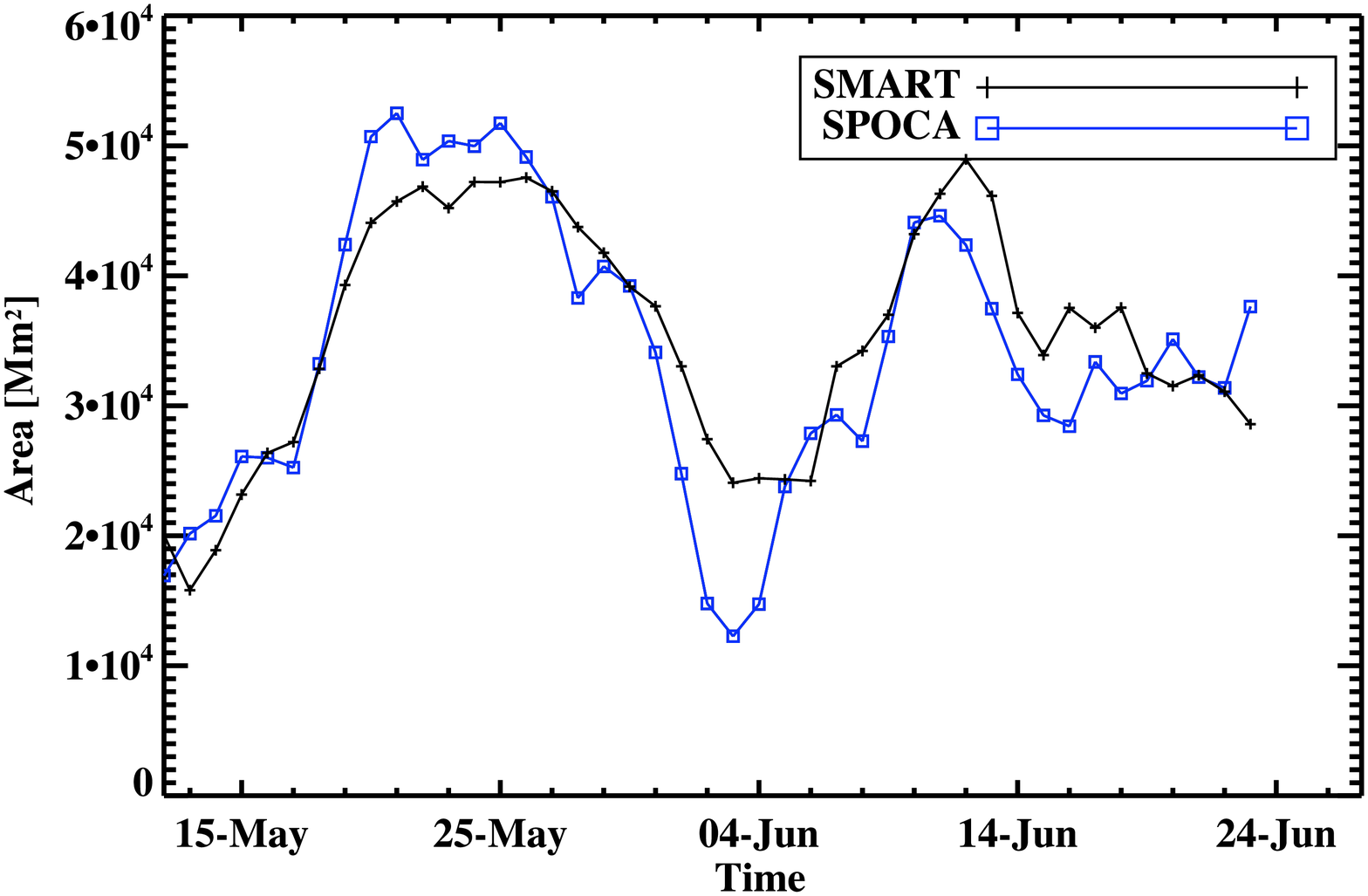}
 }
 \caption{  Comparison of average detection results of algorithms to reported
NOAA and international sunspot numbers. Upper-left: Comparison of number of
sunspots detected by \textsf{ASAP} and \textsf{STARA} and reported by NOAA and recorded
international sunspot numbers. Upper-right: Number of regions detected by \textsf{SMART} and
\textsf{SPoCA} compared with the ones reported by NOAA. Lower-left: Comparison of
average daily sunspot areas detected by \textsf{ASAP} and \textsf{STARA} \emph{versus} NOAA.
The normalised international sunspot number is over-plotted for context.
Lower-right: Comparison of average daily region areas detected by \textsf{SMART} and \textsf{SPoCA}. }
 \label{figure_performance} 
 \end{figure}

A comparison of these data is provided in Figure~\ref{figure_performance}. The
graph on the upper left side of Figure~\ref{figure_performance} compares the
daily number of sunspots detected by \textsf{ASAP} and \textsf{STARA} to the total number of spots
within NOAA regions and to the international sunspot number.
Generally, peaks and valleys in all of the series follow each
other but the international sunspot numbers and the sunspot numbers for NOAA are
usually higher than the sunspot numbers for \textsf{ASAP} and \textsf{STARA}. When sunspots are
detected manually, each umbra within a penumbra is counted as one sunspot,
whereas the automated algorithms discussed here count each penumbra as one
sunspot although it could have more than one umbra within. Therefore the
difference in sunspot numbers increases when the number of complex sunspot
regions increases. Also, the number of sunspots detected by \textsf{ASAP} is always
higher than the ones detected by \textsf{STARA}. This is because \textsf{ASAP} tends to detect
very small sunspots (sometimes pores) while \textsf{STARA} has a higher threshold for the size
of sunspot candidates.


The graph on the upper right side of Figure~\ref{figure_performance} compares
the daily number of regions detected by \textsf{SMART} and \textsf{SPoCA} to the
daily number of
NOAA regions. \textsf{SMART} and \textsf{SPoCA} detect more regions than NOAA because the NOAA
number is given to a region only if it has one or more sunspots, while \textsf{SMART} and
\textsf{SPoCA} regions do not depend on the existence of sunspots within detected
boundaries. Because of the
projection effects of large coronal loops, two close but distinct regions in the
photosphere will often be detected by \textsf{SPoCA} as one region. This explains why
\textsf{SMART} has a higher tally of daily regions than \textsf{SPoCA}. This effect is most
visible near the solar limb.

A comparison of the areas of ARs and sunspots as detected by the four algorithms
and NOAA is presented in the lower part of 
 Figure~\ref{figure_performance}. 
 \textsf{SMART}, \textsf{ASAP}, and \textsf{STARA} areas were corrected for the line-of-sight projection
effect that
 decreases the observed area as the feature moves away from the central
meridian.
 Since the line-of-sight projected area of coronal loops does not necessarily
decrease with longitude, no systematic effect is expected for the observed \textsf{SPoCA}
area, so the raw area is presented. Sunspot areas are given in millionths of
solar hemisphere to be consistent with the units of the NOAA catalogue.

The graph on the lower left side of Figure~\ref{figure_performance} compares
the sunspot areas detected daily by \textsf{ASAP} and \textsf{STARA} to the NOAA sunspot areas, while the international sunspot number is added for context.
These three time series agree well but there appears to be a one-day
shift in NOAA sunspot areas.
The \textsf{ASAP} and \textsf{STARA} sunspot measurements are averages of observations throughout the whole day, whereas, depending on the day, the NOAA sunspot observation may be quite early in the day. 
Since sunspots emerge quickly and decay slowly, any sunspots that emerge late in a day are likely to be missed by NOAA, but registered by \textsf{ASAP} and \textsf{STARA}. The following day, the new sunspots are likely to remain visible, so \textsf{ASAP} and \textsf{STARA} are likely to show the same area as the previous day, while the NOAA area will increase.

The graph on the lower right side of Figure~\ref{figure_performance} shows the
comparison between active region areas detected daily by \textsf{SMART} and by \textsf{SPoCA}.
Considering we are dealing here with photospheric \emph{versus} coronal areas, a
good agreement between the areas is obtained. Both \textsf{SMART} and
\textsf{SPoCA} areas vary smoothly. Moreover, they are large enough to include the whole
sunspot group
if present, and to measure changes in topology or complexity consistently.  
In summary, 
the time series of sunspot and AR areas are well correlated, showing similar behaviour in time, and the differences observed are
likely due to the data and detection methods used.

\subsection{Principal Component Analysis}\label{correlation}

Principal Component Analysis \citep[\textsf{PCA}:][]{Jolliffe2002} aims at reducing the dimensionality of a problem. It does so by maximizing the data structure information in the principal component space.
More precisely for a data set containing $n$ observations of $p$ variables, the principal components are the directions in
$n$-dimensional variable space in which the data set exhibits maximal variance. 

In this article, \textsf{PCA} (based on linear values) is used to get some insight in the correlation structure of the following $p$ variables:
the  Schrijver $R$ value
\citep{2007ApJ...655L.117S},
 length of the strong gradient line, magnetic flux, maximum $B$ field, area,
length of the bipole connecting line, Ising energy, and Ising energy per pixel
(Ising E ppx) as computed by the \textsf{SMART} algorithm, the sunspot area and number of sunspots as given by \textsf{ASAP}, and the raw AR area, maximum, variance, kurtosis and skewness of the EUV intensity as computed by \textsf{SPoCA}.

These variables were computed on data recorded 12 May\,--\,23 June 2003, at a cadence of 96 minutes for photospheric features, and of six hours for coronal features. Data from the various algorithms were then associated as described in Section~\ref{association_algorithm}.

We excluded data points corresponding to regions whose centre was more
than 60$^{\circ}$ from the central meridian, as projection errors involved become
too large. Table~\ref{table_PCA} lists the percentage and cumulative percentage
of the variance explained by the principal components.  The first two components explain 67\% of the total variability in the data set.

\begin{table}[hbt]
\begin{tabular}{ c c c }
Component & \% variance & Cumulative \% variance \\
1  & 52.86 &  52.86 \\
2  & 14.61 &  67.47 \\
3  & 10.30 &  77.77 \\
4  &  6.55 &  84.32 \\
5  &  5.17 &  89.49 \\
6  &  3.61 &  93.11 \\
7  &  2.82 &  95.93 \\
8  &  1.61 &  97.54 \\
9  &  0.93 &  98.47 \\
10 &  0.44 &  98.92 \\
11 &  0.39 &  99.30 \\
12 &  0.29 &  99.59 \\
13 &  0.22 &  99.81 \\
14 &  0.12 &  99.93 \\
15 &  0.07 & 100.00 \\
\end{tabular}
\caption{Percentage and cumulative percentage of the variance of the
15-dimensional variable space described above, that can be explained by the
consecutive principal components. Note that the first two principal components
comprise 67\% of the variance.}
\label{table_PCA}
\end{table}

\begin{figure}
\centerline{\hspace*{0.2\textwidth}
\includegraphics[width=0.9\textwidth,clip=]{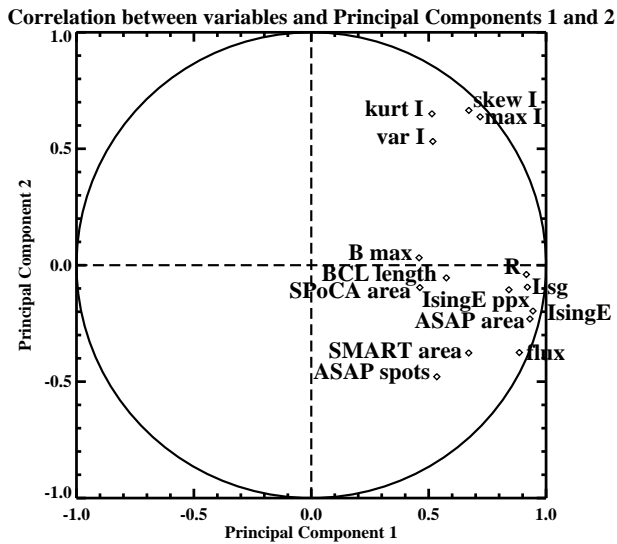}}
\caption{The projections of the algorithm variables upon the first and second
principal components are plotted. They provide a measure of the extent to which
these variables are correlated with the first and second principal components.}
\label{components_1_2}
\end{figure}

Figure~\ref{components_1_2} represents the variables in the plane of the first two components.
Each variable lies within a circle of radius one in this figure. Variables that lie close to the circle are well represented by the first two components, while variables close to the origin are not. 
The cosine of the angle formed by the origin and two points on the graph of Figure~\ref{components_1_2} gives the correlation between the two corresponding variables. 
This figure thus yields a graphical representation of the correlation structure between variables.

The first observation is that no variable is strongly anti-correlated with another in this data set. Roughly speaking, all variables evolve in the same direction.
A more detailed inspection shows that the Schrijver $R$ value, the length of the strong gradient line [Lsg], and the Ising
energy per pixel are strongly (positively) correlated to each other, 
as are the Ising Energy and the \textsf{ASAP} area. To a lesser extent, these five variables are all correlated to each other, as well as to the magnetic flux.
This related behaviour of PSL length, $R$ value, Ising energy, and \textsf{ASAP} area is
apparent in Figures \ref{10377evolve_area_pos_num},
\ref{10377evolve_complexity}, \ref{10365evolve_area_pos_num},
\ref{10365evolve_complexity}, \ref{10365_2evolve_area_pos_num} and
\ref{10365_2evolve_complexity}.
These variables are linked to a measure of complexity of the AR and its
capability to produce a flare, see \citet{Colak2010a}.

The maximum and skewness of the EUV intensity are strongly correlated: indeed a high value of maximum EUV intensity implies a long tail in the intensity distribution function, hence a high skewness. Similarly, the variance and kurtosis of the EUV intensity are strongly correlated, which is expected since they are related by definition. Note that variance and kurtosis lie further from the circle than maximum and skewness, and hence are less well represented by the first two principal components.
Finally, the maximum magnetic field, length of the Bipole Connecting Line, \textsf{SPoCA} area, \textsf{SMART} area, and \textsf{ASAP} number of sunspots are not well represented by the two first components, and hence their correlation structure cannot be interpreted from this plot.

\textsf{PCA} also tends to separate photospheric and coronal contribution. Features computed at the photospheric level such as $R$, Lsg, Ising energy, \textsf{ASAP} area, and flux have a large contribution to the first component, which accounts for 52.86\% of the variability in the total data set.
The maximum, variance, skewness, and kurtosis in EUV intensity images are moderately correlated to both the first and second principal components.

This study shows that a reduction in dimensionality using \textsf{PCA} can be performed without losing too much information.  Such reduction can enhance the accuracy and robustness of a subsequent classification scheme~\citep{jiang2011} that would aim for example at separating active regions that are prone to flares from quiet active regions.

\subsection{Case Studies}\label{case_studies}

In the following section we analyse the time evolution of the ARs that emerge
as NOAA 10377 (a simple region) and 10365 (a complex, flaring region). Of special interest is how activity in the corona
results from changes in the photosphere. Drawing this connection is essential
for flare prediction, since the photosphere is more easily physically
characterised than the corona, where flares actually occur. 
The photosphere--corona connection
is not well understood, \emph{e.g.} the work of \citet{2007ApJ...656.1173L} and of
\citet{Handy2001}, with the references therein.

We compare 
observations of NOAA 10365 and 10377 with flares characterised
by the
\emph{Reuven Ramaty High Energy Solar Spectroscopic Imager} (RHESSI) team and
distributed in the RHESSI flare
list( \url{http://sprg.ssl.berkeley.edu/~jimm/hessi/hsi_flare_list.html}).
The flares, which have been associated with the individual ARs by the RHESSI
team, are represented in plots (Section \ref{case_studies}) as downward pointing
arrows, whose size is logarithmically proportional to their peak count rate.

\subsubsection{NOAA 10377}\label{noaa_10377}

\begin{figure}
\centerline{\includegraphics[width=0.9\textwidth,clip=]{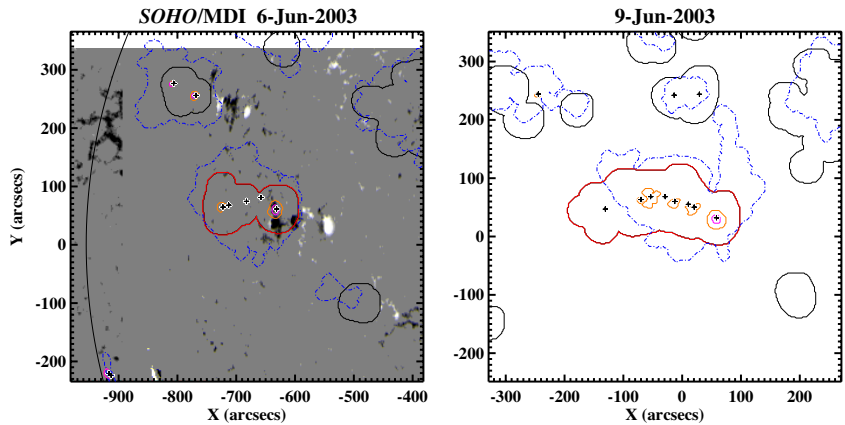}}
\caption{A comparison of NOAA AR 10377 detections. \textsf{ASAP} sunspots are
represented by black crosses. The contours represent \textsf{SMART} in black (with NOAA
10377 outlined in red) for the magnetic features, \textsf{SPoCA} in dashed blue for
coronal features, and \textsf{STARA} in orange for sunspot penumbrae and magenta for
umbrae.}
\label{detection10377compare}
\end{figure}

NOAA 10377 first emerges just before rotating onto the visible disk on 4 June
2003. It continues to gradually develop as it progresses across the disk
producing very little activity (only one B9.1 event is listed in the NOAA events
catalogue ({\url{http://www.swpc.noaa.gov/ftpdir/indices/events/README})).
Some of the flares produced by 10377 may have been missed due to the
presence of 10375, which produced many large flares, swamping any signal that
could be attributed to 10377.

Figure \ref{detection10377compare} shows the \textsf{SMART} detection of 10377 in red,
while other features are outlined in black. The extended dashed blue contours
are \textsf{SPoCA} AR detections and the small symbols and contours are sunspot
detections from \textsf{ASAP} and \textsf{STARA}, respectively. It is clear from Figure
\ref{detection10377compare} that positions of the \textsf{SMART}, \textsf{ASAP}, \textsf{STARA}, and \textsf{SPoCA}
detections agree quite well. Whereas the sunspots detected by \textsf{ASAP} and \textsf{STARA} are
well confined within the \textsf{SMART} magnetic region boundary, the \textsf{SPoCA} region most
often contains most of the \textsf{SMART} detection. In the case of coronal loop structures forming between nearby
ARs, adjacent \textsf{SPoCA} detections will merge and the \textsf{SMART} and \textsf{SPoCA}
centroids will diverge. This is especially apparent near the solar limb, where
coronal structures extending above the solar surface will be
superimposed.

\begin{figure}
\centerline{\includegraphics[width=0.9\textwidth,clip=]{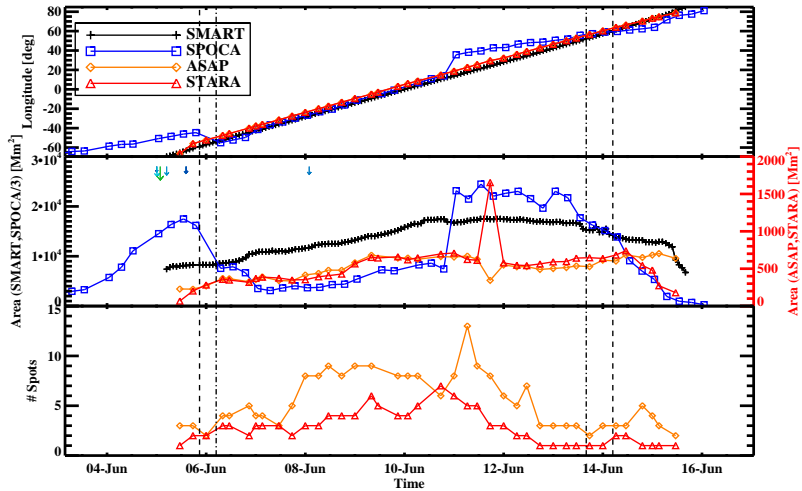}}
\caption{Time series of position, area, and sunspot information characterising
the evolution of NOAA AR 10377. The legend indicates symbols and colors for each
of the detection algorithms. The axes of the area plot are split between left
(\textsf{SPoCA} and \textsf{SMART}) and right (\textsf{ASAP} and \textsf{STARA}). The \textsf{SPoCA} areas have been divided
by three for display.}\label{10377evolve_area_pos_num}
\end{figure}

Figures~\ref{10377evolve_area_pos_num}\,--\,\ref{10377evolve_complexity} show the
evolution of NOAA 10377 as it progresses across
the disk. In the top panel of Figure \ref{10377evolve_area_pos_num}
the Stonyhurst longitudes of the region centroids from each algorithm are shown.
The vertical dotted lines indicate where the AR magnetic bounding box edges
(dashed--dotted) and centroid (dashed) cross $-60$ and 60$^{\circ}$ longitude. The
cosine correction used to correct for line-of-sight effects on magnetic-field
properties is not sufficient outside of
this range. Also, beyond 60$^{\circ}$, sunspot visibility is below
$\approx\frac{1}{3}$ of that at
disk centre \citep{Watson2009} due to the Wilson depression.

The top panel of Figure~\ref{10377evolve_area_pos_num} tracks the longitude of centroids over time. We see that \textsf{ASAP} and \textsf{STARA} curves are above the \textsf{SMART} curve on this plot, suggesting that  the centroid of the magnetic footpoints (\textsf{SMART}) follows behind the sunspot
centroids (\textsf{ASAP} and \textsf{STARA}).
Since the longitudinal speed of the white-light and
magnetic detections are the same, this implies that the following polarity of the
AR extends beyond the embedded sunspots, while the leading polarity remains
compact. As the NOAA region 10377 is close to 10375, this last region affects the \textsf{SPoCA}
detections. From 3\,--\,6 June, \textsf{SPoCA} detects both NOAA regions
within a single boundary. When this region splits into two parts on 6 June, the
\textsf{SPoCA} longitude and area curves decrease abruptly, and can now be directly
compared to the photospheric structures. This changes when the two NOAA regions
merge again on 11 June. Whenever the region detected by
\textsf{SPoCA} corresponds to the region detected by the other algorithms, all four
longitudes agree well.

The total sunspot area determined by \textsf{ASAP} and \textsf{STARA} (Figure \ref{10377evolve_area_pos_num}, middle) is very similar except for one data point near 12 June 2003. This is due to the MDI image on 11 June 2003 at
1736 UT being distorted. Most of the distortion is visible on
the south limb of the image where this area is darker than the rest of the solar
disk. Because \textsf{ASAP} detects the solar disk directly from the image, while \textsf{STARA}
uses FITS keywords, the determination of the solar disk by these two methods is
different. This explains why on this image the \textsf{ASAP} sunspot area is much smaller
than the \textsf{STARA} area: whereas the distorted area is detected by \textsf{STARA} as a
large sunspot, it is completely discarded by \textsf{ASAP}.
The \textsf{SMART} and \textsf{SPoCA} areas of photospheric magnetic regions and coronal active
regions obey the same general trend as the sunspot areas, although the absolute
scales are different. While the area measurements are stable, the total
number of sunspots is not. The total area is dominated by the largest
sunspots, while the total number of spots is affected by small transients
which \textsf{ASAP} is especially sensitive to.

\begin{figure}
\centerline{\includegraphics[width=0.9\textwidth,clip=]{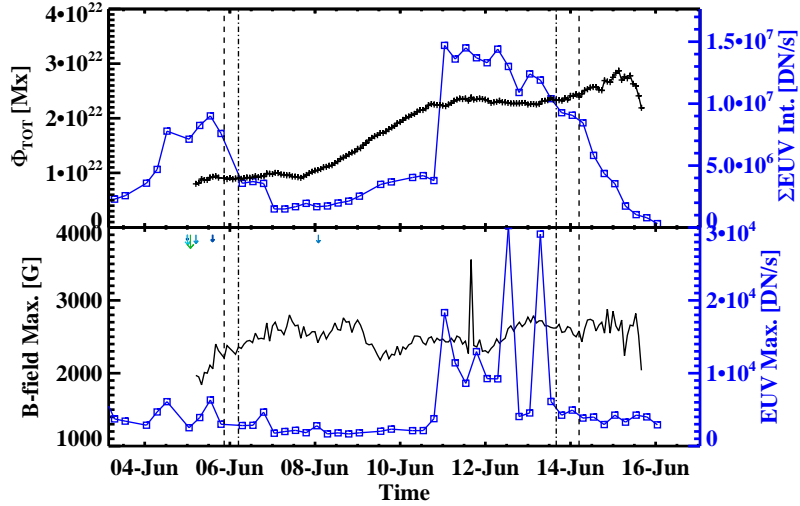}}
\caption{Time series of (top) total magnetic flux, total EUV intensity, (bottom)
maximum magnetic field, and maximum EUV intensity for NOAA AR 10377. The axes of
the plots are split between left (magnetic-field properties, black crosses) and
right (coronal properties, blue squares). RHESSI flares associated with the AR
are indicated by downward arrows.}\label{10377evolve_flux_bmax}
\end{figure}

In the top panel of Figure \ref{10377evolve_flux_bmax}, the emergence of the
magnetic structure of 10377 is clearly seen in measurements of its total flux. 
The AR is stable until $\approx$\,8 June 2003 when a phase of rapid emergence begins,
lasting until $\approx$\,11 June when the total magnetic flux has more than
doubled.
Comparing Figure
\ref{10377evolve_area_pos_num} and Figure \ref{10377evolve_flux_bmax}, we see
that the total magnetic flux increases faster than the magnetic area, implying
that the AR magnetic fields emerge relatively faster than they
diffuse. The same general smooth trend is observed in the \textsf{SPoCA} total EUV intensity
between 6 and 11 June. After NOAA 10377 merges with 10375 on 11 June, we see a
clear decay of the total EUV intensity in
this combined region. Note that both \textsf{SMART} flux and \textsf{SPoCA} total EUV
intensity behave similarly to the region area time series.

In the bottom panel the maximum magnetic
field is much less stable than the flux, and shows no clear trend.
The maximum \textsf{SPoCA} EUV intensity does not change significantly between 6 and 11
June for
NOAA region 10377, but exhibits three clear peaks afterwards which can be attributed to region 10375.
The first peak, on 11 June, can be attributed to \textsf{SPoCA} merging with NOAA 10375.
The peak on 12 June, near 1300, is probably associated with the M1.0
flare in 10375 at around 1358\,UT, whereas the 13 June 0700\,UT peak
is probably related to the M1.8 (0628\,UT) or C6.1 (0710\,UT) flares in 10375. These flares (appearing in the NOAA events catalogue) are not indicated
by the RHESSI arrows, since we have only displayed those flares attributed to 10377. 
This shows that \textsf{SPoCA} maximum intensity is capable of indicating solar eruptions.

\begin{figure}
\centerline{\includegraphics[width=0.9\textwidth,clip=]{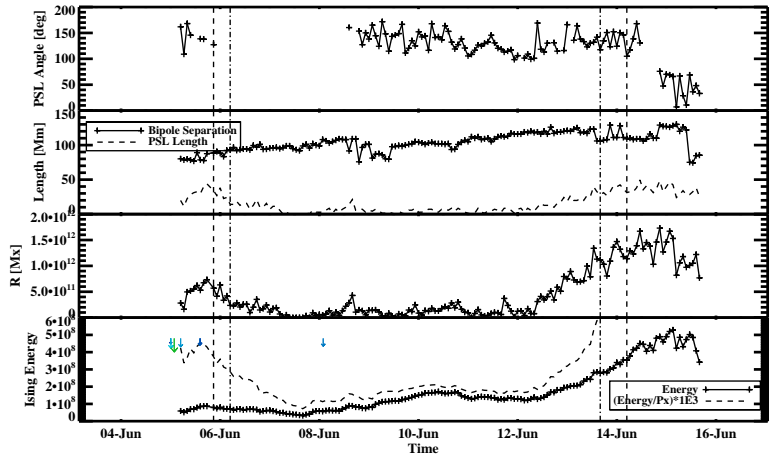}}
\caption{Time series of (top) PSL orientation with respect to the bipole
separation line, (middle-top) bipole separation line length (crosses) and PSL
length (dashed), (middle-bottom)  $R$, and (bottom) Ising energy (crosses) and
Ising energy per pixel (dashed; multiplied by 1000 for display) for NOAA AR
10377.}\label{10377evolve_complexity}
\end{figure}

Magnetic properties related to polarity mixing and complexity are shown in
Figure \ref{10377evolve_complexity}. In the top panel, the angle between the
bipole connection line and polarity separation line (PSL) is presented.  Since 
the PSL in this AR is only  a few megameters (or pixels) long (\emph{cf.} middle-top panel), this angle cannot be measured in a reliable way. Indeed,  a small growth in the PSL detection in any direction can cause the angle to change dramatically.
In the middle-bottom panel the total
flux near the PSL [$R$] is very small until it begins to increase as a false PSL
is detected due to the near-horizontal fields of the large leading polarity
sunspot approaching the west limb on 12 June.

Ising energy, a proxy for magnetic connectivity, is shown in the bottom panel.
This property increases during the main magnetic emergence phase ($\approx$\,8 to 10
June 2003) since it is dependent on the magnetic-field strength and inversely
dependent on the distance between individual magnetic elements. The Ising energy
per pixel (dashed line) appears to be very susceptible
to geometrical effects as the large decrease near the west limb and increase
near the east limb both coincide with the formation of false PSLs in the leading
sunspot. It should be noted that this quantity was calculated without remapping
the data to disk-centre as done by \citet{Ahmed2010}, giving the measurement an even
larger viewing-angle dependence.

\subsubsection{NOAA 10365}\label{noaa_10365}

\begin{figure}
\centerline{\includegraphics[width=0.9\textwidth,clip=]{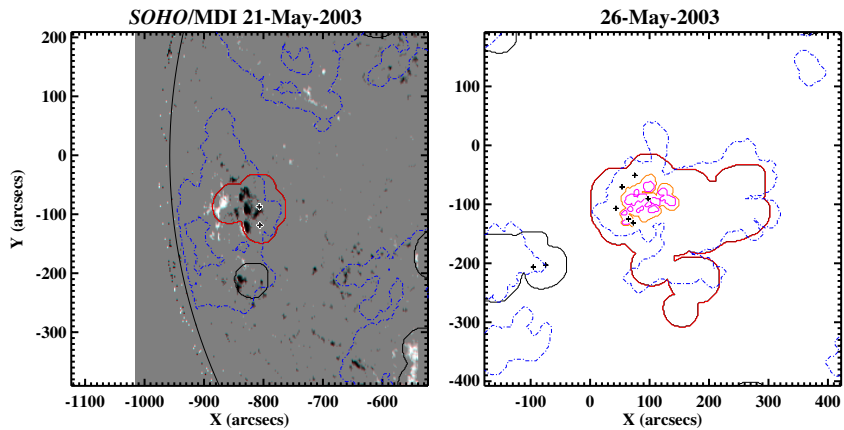}}
\caption{A comparison of detection contours for NOAA AR 10365. \textsf{ASAP} sunspots are
represented by black crosses. The contours represent \textsf{SMART} in black (with NOAA
10365 outlined in red) for the magnetic features, \textsf{SPoCA} in dashed blue for
coronal features, and \textsf{STARA} in orange for sunspot penumbrae and magenta for
umbrae.}
\label{detection10365compare}
\end{figure}

Active region NOAA 10365 rotates onto the visible solar disk on 19 May 2003 at
heliographic latitude -5$^{\circ}$. At this point 10365 is mature and decaying,
having emerged and evolved on the far side of the Sun. On 24 May, a new
bipolar structure rapidly emerges in the extended plage of the trailing
(positive) polarity. NOAA switches the 10365 designation to this newly emerged
bipole several days later. As the bipole evolves it develops a strong double PSL
by merging with the decayed flux. It produces many C- and M-class flares
and several X-class flares. The AR progresses around the visible disk,
eventually returning as NOAA 10386. The onset of decay occurs as C- and M-class
flares are produced with decreasing frequency and the spot areas, magnetic flux,
and field strengths decrease.

Figure \ref{detection10365compare} shows a comparison of the heliographic
positions and sizes of two sets of \textsf{SMART}, \textsf{ASAP}, \textsf{STARA} and \textsf{SPoCA} detections of
NOAA 10365. We can see that positions of the \textsf{SMART}, \textsf{ASAP}, \textsf{STARA} and \textsf{SPoCA}
detections agree well. The \textsf{SPoCA} detection, however, includes coronal
loops extending away from the footpoint boundary of NOAA 10365.
Before 24 May, \textsf{SPoCA} merges NOAA region 10367 with its detection of 10365. From
24
May\,--\,27 May it only detects 10365, on 27 May at 1300\,UT there is a
single data point where these regions are merged by \textsf{SPoCA}, and from 29 May at
0100\,UT onwards, \textsf{SPoCA} merges them for the remaining observation period. The
longitudes of all
detections within 24\,--\,27 May agree well. After 27 May the \textsf{SPoCA} longitude
drifts, reflecting changes in the merged coronal structures.
Unlike 10377, the magnetic centroid of 10365 at first trails behind the sunspot
centroid but then precedes it, as evidenced by the top panel in Figure
\ref{10365evolve_area_pos_num}. This is because the new bipole, which develops
many spots, emerges behind the existing weakly spotted bipole. The new emergence
is clear in the plot of total sunspot area (middle panel), and is unclear in the
magnetic and EUV area plots since the new bipole emerges partially within the
boundary of the old one. Note that all areas for NOAA 10365 are much larger than
those for simpler region 10377. For \textsf{SMART}, area is very sensitive to weak
magnetic plage, however. This can be seen in the sudden jumps around May 25 and
28, which are due to nearby plage temporarily merging with the AR.   
The jump in \textsf{STARA}  area on 27 May can be attributed to a bad data file (note that there is no \textsf{ASAP}
data point at that time).

\begin{figure}
\centerline{\includegraphics[width=0.9\textwidth,clip=]{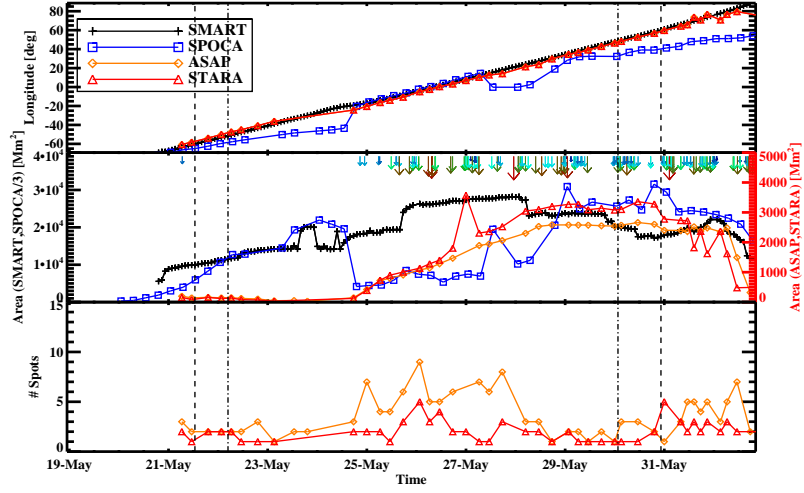}}
\caption{Time series of position, area, and sunspot information characterising
the evolution of NOAA AR 10365. The legend indicates symbols and colors for each
of the detection algorithms. The axes of the area plot are split between left
(\textsf{SPoCA} and \textsf{SMART}) and right (\textsf{ASAP} and \textsf{STARA}). The \textsf{SPoCA} areas have been divided
by three for display.}\label{10365evolve_area_pos_num}
\end{figure}

\begin{figure}
\centerline{\includegraphics[width=0.9\textwidth,clip=]{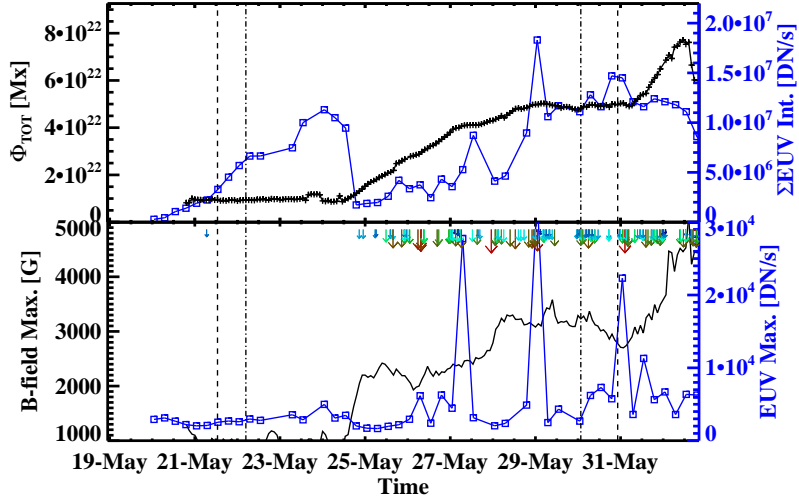}}
\caption{Time series of (top) total magnetic flux, total EUV intensity, (bottom)
maximum magnetic field, and maximum EUV intensity for NOAA AR 10365. The axes of
the plots are split between left (magnetic-field properties, black crosses) and
right (coronal properties, blue squares). RHESSI flares associated with the AR
are indicated by downward arrows.}\label{10365evolve_flux_bmax}
\end{figure}

From 25 May onwards, the total magnetic flux increases gradually to over
four-fold the initial value during development and levels off around 29 May (see
Figure~\ref{10365evolve_flux_bmax}). 
The maximum magnetic field increases
abruptly on 25 May and also increases over time, albeit less smoothly
than the magnetic flux. The maximum magnetic field did not show an
overall increasing or decreasing trend in the case of simpler NOAA region
10377.

The time series of \textsf{SPoCA} maximum intensity exhibits some peaks, which can be related to the following flares produced by NOAA 10365: the M1.9 flare at 0534\,UT on 26 May, the
M1.6 flare at 0506\,UT on 27 May, the X1.2 flare  at 0051\,UT
on 29 May, and the M9.3 flare at 0213\,UT on 31 May.
The last two flares are even visible in the total \textsf{SPoCA} intensity, which shows
more or less a gradual increase over time, but less smooth than in the case of
the simpler NOAA region 10377. The flares not picked up by \textsf{SPoCA} likely occurred
in between EIT images. 

\begin{figure}
\centerline{\includegraphics[width=0.9\textwidth,clip=]{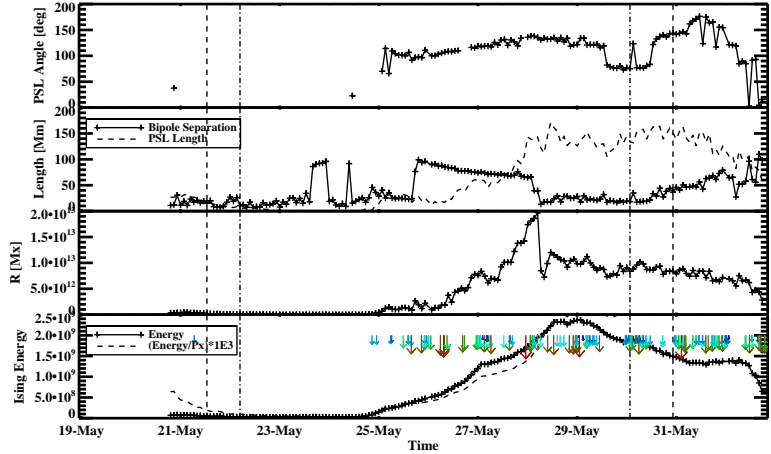}}
\caption{Time series showing proxies for the complexity and polarity mixing in NOAA AR
10365. (middle-top) bipole separation line length (crosses) and PSL
length (dashed), (middle-bottom)  $R$, and (bottom) Ising energy (crosses) and
Ising energy per pixel (dashed; multiplied by 1000 for display)}\label{10365evolve_complexity}
\end{figure}

Signatures in the evolution of the magnetic topology of NOAA 10365 precede its
intense coronal activity, indicated by the associated RHESSI flares in Figure
\ref{10365evolve_complexity}. Just before 25 May
2003 the new emergence causes a jump in the main bipole separation line length.
As the emergence continues and strong PSLs develop, this length decreases,
while the total PSL length increases, as shown in the middle-top panel of Figure
\ref{10365evolve_complexity}. Also, there are signs of gradual helicity
injection
as the angle between the main bipole connection line and the main PSL grows from
near perpendicular ($90^{\circ}$) to around $120^{\circ}$ (top panel). The flux
near PSL [$R$]
grows during this time, as does the Ising energy (middle-bottom and bottom
panels, respectively). A bump in $R$ just before 26 May is followed by an
intense
RHESSI flare. Intense flaring begins again around the second bump in $R$ on 28
June. Examining the development of Ising energy, it appears that sharp increases in the property
are followed by the most
intense flaring.


\begin{figure}
\centerline{\includegraphics[width=0.9\textwidth,clip=]{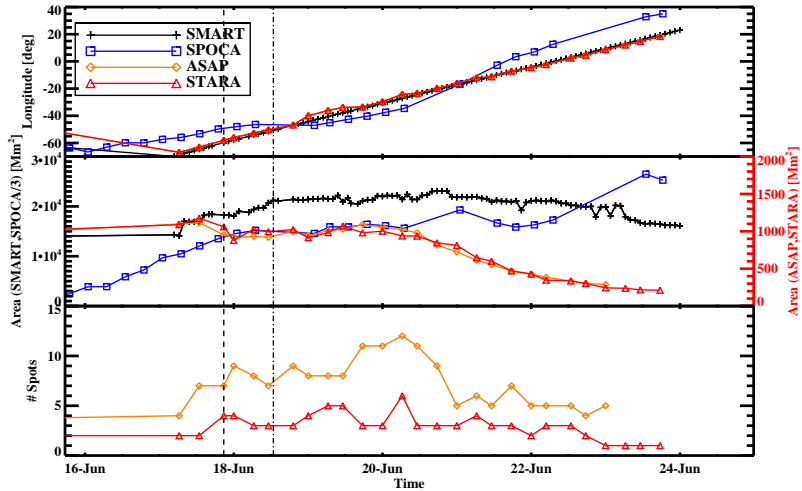}}
\caption{Time series of position, area, and sunspot information characterising
the decay phase of NOAA AR 10365 (renamed 10386) during its second disk passage.
The legend indicates symbols and colors for each
of the detection algorithms. The axes of the area plot are split between left
(\textsf{SPoCA} and \textsf{SMART}) and right (\textsf{ASAP} and \textsf{STARA}). The \textsf{SPoCA} areas have been divided
by three for display.}\label{10365_2evolve_area_pos_num}
\end{figure}

NOAA 10365 returns for a second disk passage, renamed 10386. We are able to
observe its decay phase, as shown in Figures \ref{10365_2evolve_area_pos_num}\,--\,\ref{10365_2evolve_complexity}. As
no RHESSI data on flares is available for this period, no flare arrows were
added to these figures.
While the longitude of the \textsf{SMART} magnetic centroid increases linearly with
time, the \textsf{ASAP} and \textsf{STARA} sunspot centroids show small departures from this line
between 19 and 21 June, preceding the magnetic centroid.
The \textsf{SPoCA} detection of NOAA 10386 merges with
10388 and 10389, so a direct comparison with the other algorithms cannot be
made.

\begin{figure}
\centerline{\includegraphics[width=0.9\textwidth,clip=]{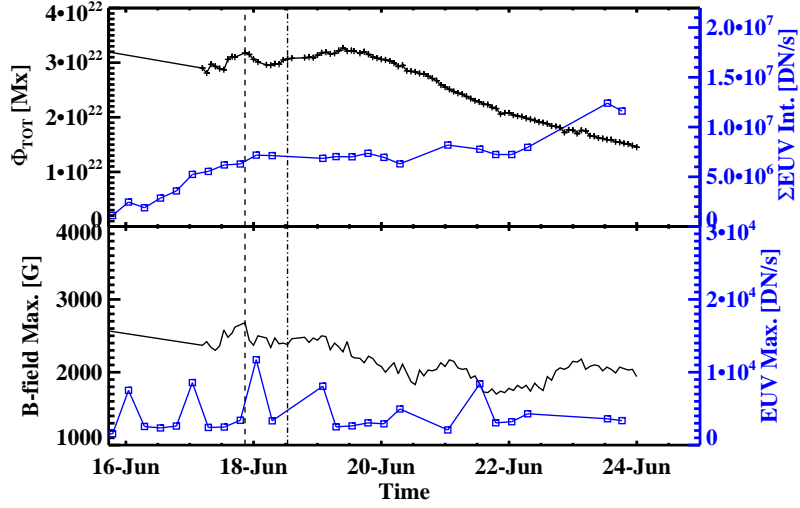}}
\caption{Time series of (top) total magnetic flux, total EUV intensity, (bottom)
maximum magnetic field, and maximum EUV intensity for NOAA AR 10365 on its
second disk passage as 10386. The axes of
the plots are split between left (magnetic-field properties, black crosses) and
right (coronal properties, blue squares).}\label{10365_2evolve_flux_bmax}
\end{figure}

\begin{figure}
\centerline{\includegraphics[width=0.9\textwidth,clip=]{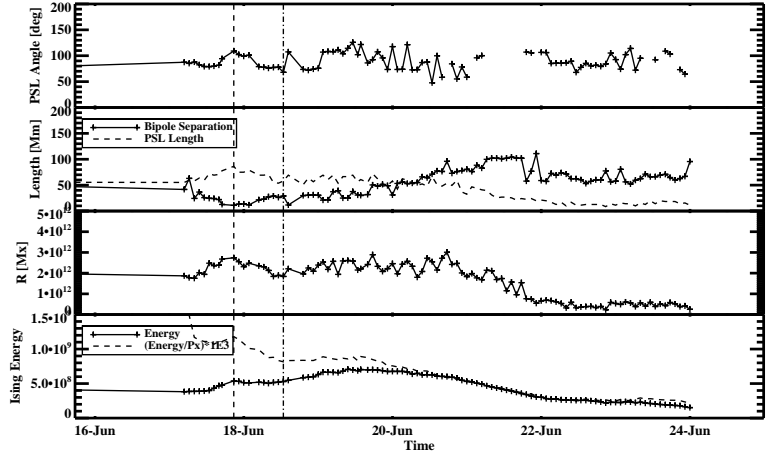}}
\caption{Time series showing proxies for the complexity and polarity mixing in
NOAA 10386. (top) PSL orientation with respect to the bipole
separation line, (middle-top) bipole separation line length (crosses) and PSL
length (dashed), (middle-bottom)  $R$, and (bottom) Ising energy (crosses) and
Ising energy per pixel (dashed; multiplied by 1000 for display)}\label{10365_2evolve_complexity}
\end{figure}

The magnetic area does not change significantly, but the total sunspot area
clearly
decreases (middle panel, Figure~\ref{10365_2evolve_area_pos_num}), and has already decreased substantially since the
previous
disk passage (as NOAA 10365).
The total magnetic flux decreases (top panel, Figure
\ref{10365_2evolve_flux_bmax}) as its magnetic fields diffuse and weaken.
Comparing the values to Figure~\ref{10365evolve_flux_bmax}, we notice that the
flux had already decreased significantly since the previous solar rotation.
The total EUV intensity does not change substantially, regardless of the
weakening magnetic footpoints, although it has decreased since the previous
solar rotation. Its increase on 22 June is due to the detection
merging with a large region near the limb.
The maximum magnetic-field value shows a gradual, although not very smooth
decrease, and has also decreased since the previous passage.
The maximum EUV intensity does not show a clear trend, and although several jumps
are detected, the intensity levels are much less than those associated with flares
over the previous passage. The peak on 18 June at 0100\,UT, for instance,
can likely be associated to the M6.8 flare produced by region 10386 at
2227\,UT on 17 June.
The PSL length has decreased since the previous solar rotation, and shows a
further gradual decrease in Figure~\ref{10365_2evolve_complexity}.
The same is true for both $R$ and the Ising energy.

\section{Discussion and Conclusion}\label{disc_concl}

In this article the performance of the presented algorithms is tested in several ways: the overall detection performances are investigated as compared to established methods; the correlations between extracted physical properties are determined using Principal Component Analysis; the stability and usefulness of the algorithms and of various feature properties for studying feature time-evolution is tested. Additionally, some lessons were learned about data analysis using automated detection algorithms. The following relates the main results of these investigations.


In Section \ref{performance} we compared \textsf{ASAP}, \textsf{STARA}, SIDC, and NOAA sunspot detections and found out that although different numbers of features are detected, the total disk areas agree well. The same is true for the comparison between \textsf{SMART} and \textsf{SPoCA} active-region detections. The difference in numbers can be explained as follows: \textsf{ASAP} is sensitive to pores while \textsf{STARA} only detects developed sunspots. Pores clearly do not add significant area to the total disk coverage, while there is a large number of them detected. NOAA sunspot counts include both penumbrae and umbral cores, whereas \textsf{ASAP} and \textsf{STARA} only count sunspot groups. Consequently, sunspot counts should be very carefully assessed when applied to a long-term inhomogeneous historical record. NOAA only counts those active regions that possess sunspots, which is not the case for \textsf{SMART} nor \textsf{SPoCA}. \textsf{SMART} detects both small AR fragments as well as complexes of bipolar structures, while \textsf{SPoCA} is sensitive to bright loop connections between adjacent features, often merging them into a single detection. These differences are partially dependent on thresholds chosen by the developers and may be tuned to return an agreeable feature boundary. Boundaries are inherently arbitrary as there is no established definition of each feature. Additionally, features evolve continuously and are prone to merging and fragmentation, so universally defining a feature is very difficult.  The summed area of features is a preferred quantity as it is both more stable and less ambiguous.

The use of Principal Component Analysis (Section \ref{correlation}) has allowed us to determine which feature properties contain the most information about our data set. \textsf{PCA} also tends to separate photospheric and coronal contribution.
Features computed at the photospheric level such as $R$, Lsg, Ising energy, \textsf{ASAP} sunspot area, and magnetic flux are highly correlated to each other and have a large contribution to the first component, which accounts for 52.86\% of the variability in the total data set.
The maximum, variance, skewness, and kurtosis in EUV intensity images are highly correlated to each other and moderately correlated to both the first and second principal components.
By reducing dimensionality the accuracy and robustness of a classification scheme can be enhanced~\citep{jiang2011}. For example, this could be used to discriminate between flaring and non-flaring active regions properties.

Through the time series analysis of two AR case studies (Section
\ref{case_studies}), we have observed three physical processes evident in their
evolution: emergence of a bipolar magnetic structure, sunspots, and EUV loops
(Section \ref{noaa_10377}); increase and peak in non-potentiality,
followed by the onset of flaring (Section \ref{noaa_10365}); decay and
weakening of magnetic footpoints (Section \ref{noaa_10365}).
We find that the algorithms show good correspondence between centroid positions and areas but significant divergence is seen in other properties. 

In the case study of the simple active region NOAA 10377 (Section \ref{noaa_10377}) we see that the total number of detected sunspots fluctuates wildly as transient spots rapidly emerge and disappear. This is partly due to the visibility curve since the area of small spots is highly impacted by the observer's viewing angle \citep{Dalla2008,Watson2009}. 
In order to study AR evolution, sunspot area is more indicative of the emergence and decay of an AR
than coronal or magnetic area, which do not necessarily decrease during decay (see Section \ref{case_studies}). As a basis for long-term AR tracking, magnetic flux is more useful than the sunspot area (since sunspots are much more transient than their magnetic footprints) or maximum magnetic field value (since it is affected by the MDI saturation problem
\citet{liu_etal_2007}).
Also, the maximum magnetic-field value is unstable since different positions in the active region will over-take each other in field magnitude as they develop, causing the location that the value is sampled from to vary wildly. Finally, the total EUV intensity determined by \textsf{SPoCA} has a smooth behaviour over time and is
closely linked to area. The maximum EUV intensity peaks when the active region emits a large flare, and appears to be a useful indicator of eruptions in the corona.

The case study of complex, flaring NOAA 10365 (Section \ref{noaa_10365}) shows that flaring can happen both in periods of flux emergence as well as non-potentiality enhancement in an active region. Following the initial flaring during the emergence phase of evolution, further flaring occurs as the main PSL rotates with respect to
the bipole connection line. This is a sign of helicity injection and is coincident with increases of other properties related to polarity mixing. Helicity injection has been established as a method of increasing non-potentiality and may be caused by the emergence of subsurface twisted flux ropes, as seen in \citet{2007ApJ...657..577D}.

As NOAA 10365 returns after one solar rotation, decay is seen in the strength of its magnetic
footprint. However, the area is not seen to decrease significantly, since supergranular diffusion causes a radial dispersal of magnetic elements. Coronal structures do not appear to decay readily, either. This result agrees with \cite{Lites1995}, where it is reasoned that if the coronal magnetic structure is closed, it may be in a state of quasi-static equilibrium, whereby the magnetic buoyancy of the loops is cancelled by the weight of plasma trapped at the bottom of the closed structure.

By performing these studies, it is found that magnetic active-region detections provide the most stable base for feature tracking. Sunspots are only visible for short periods of time, and coronal detections continually form bright loop connections to nearby features. The simple feature tracking method used in this article (see Section \ref{association_algorithm}) is novel in that it allows features to be tracked between multiple disk passages. This is essential for analysing the complete life-cycle
of an active region, as exemplified in the analysis of NOAA 10365 (Section \ref{noaa_10365}). Future work on active-region evolution should combine morphological information to better handle merging and splitting (as done by \citet{WelschLongcope2003}) with our method of multiple disk passage tracking.
Future work on our algorithms will also address the automatic detection and handling of structural or visible errors in solar data, to avoid discontinuities in the time series due to a corrupted image, as was seen in the \textsf{STARA} outputs used in the case studies (see Section \ref{noaa_10377}).

This work will be expanded in the future to include an analysis of the full SOHO archive as well as detailed studies of photospheric and coronal SDO data sets. Many physical studies will benefit from this work, as investigations that examine coronal heating as a result of large scale magnetic fields \citep{Schrijver1987,Fisher1998}, coupling between the photosphere and corona \citep{Handy2001}, sources of coronal mass ejections \citep{Subramanian2001}, flux emergence and distribution \citep{Liu2004,Abramenko2005} and
flare forecasting \citep{Gallagher2002} can all be repeated with these automated
detection methods. Using these methods allows a far greater number of features
to be analysed and reduces human bias in the detection of features in
the solar data.

The algorithms presented here are automated (once thresholds have been fixed), independent, and unsupervised. Although some development remains to be done, they detect features the way that they are intended, and will provide useful additions to the SDO pipeline feature-detection methods. However, this work shows that automated methods cannot replace human data analysis but they can help to stream-line the process.

%

%

%

%
\begin{acks}
Funding of CV and VD by the Belgian Federal Science Policy Office (BELSPO) through the
ESA/PRODEX SIDC Data Exploitation program, as well as by the Solar--Terrestrial
Center of Excellence/ROB, is hereby appreciatively acknowledged. FTW acknowledges
Ph. D. funding from the Science and Technology Facilities Council and the
guidance of his supervisor, Lyndsay Fletcher. We acknowledge
support from ISSI through funding for the International Team on ``Mining and
exploiting SDO data in Europe'' led by V. Delouille. \textsf{ASAP} is supported by an EPSRC Grant
(EP/F022948/1), which is entitled ''Image Processing, Machine Learning and
Geometric Modelling for the 3D Representation of Solar Features''. PAH
acknowledges support from 
ESA/PRODEX and a grant from the EC Framework Programme 7 (HELIO) and the
guidance of his supervisor, Peter T. Gallagher. We would like to thank the SOHO
team for making both their data and analysis software publicly available, Omar
W. Ahmed for use of the Ising Energy software, and Aidan M. O'Flannagain for
advice on the RHESSI flare list. 
\end{acks}

%
%
 \bibliographystyle{spr-mp-sola}
 \bibliography{issi_database}  
%
%
%
%

} 

\end{article} 
\end{document}